\documentclass[twocolumn]{aastex63}
\pdfoutput=1 
\usepackage[T1]{fontenc}
\usepackage{amsmath,amstext}
\usepackage{apjfonts} 
\usepackage[figure,figure*]{hypcap}
\usepackage{microtype}
\usepackage{tablefootnote}
\usepackage{booktabs}  
\usepackage{longtable}
\usepackage{hyperref}

\begin{document}
\shortauthors{Eftekhari et al.}

\shorttitle{AT2022cmc: The Relativistic Jet Shuts Off}

\title{Late-time X-ray Observations of the Jetted Tidal Disruption Event AT2022cmc: The Relativistic Jet Shuts Off}
\newcommand{\NU}{\affiliation{Center for Interdisciplinary Exploration and Research in Astrophysics (CIERA) and Department of Physics and Astronomy, Northwestern University, Evanston, IL 60208, USA}}

\newcommand{\UA}{\affiliation{Department of Astronomy and Steward Observatory, University of Arizona, 933 North Cherry Avenue, Tucson, AZ 85721-0065, USA}}

\newcommand{\UCB}{\affiliation{Department of Astronomy, University of California, Berkeley, CA 94720-3411, USA}}

\newcommand{\UTAH}{\affiliation{Department of Physics \& Astronomy, University of Utah, Salt Lake City, UT 84112, USA}}

\newcommand{\UCBphys}{\affiliation{Department of Physics, University of California, 366 Physics North MC 7300, Berkeley, CA 94720, USA
}}

\newcommand{\STSCI}{\affiliation{Space Telescope Science Institute, 3700 San Martin Dr, Baltimore, MD 21218, USA}}

\newcommand{\DARK}
{\affiliation{DARK, Niels Bohr Institute, University of Copenhagen, Niels Bohr
Building (NBB), Jagtvej 155A, 1. floor, 2200 Copenhagen N.,
Denmark}}

\newcommand{\CfA}{\affiliation{Center for Astrophysics ${\rm \mid}$ Harvard {\rm \&} Smithsonian, 60 Garden St, Cambridge, MA 02138, USA}}

\newcommand{\OR}{\affiliation{Department of Physics, University of Oregon, Eugene, OR 97403, USA}}

\newcommand{\MIT}{\affiliation{Kavli Institute for Astrophysics and Space Research, Massachusetts Institute of Technology, Cambridge, MA 02139, USA}}

\newcommand{\Miller}{\affiliation{Miller Institute for Basic Research in Science, 468 Donner Lab, Berkeley, CA 94720, USA}}
\author[0000-0003-0307-9984]{T. Eftekhari}
\altaffiliation{NHFP Einstein Fellow}
\NU
\author[0000-0002-9182-2047]{A. Tchekhovskoy}
\NU
\author[0000-0002-8297-2473]{K. D. Alexander}
\UA
\author[0000-0002-9392-9681]{E. Berger}
\CfA
\author[0000-0002-7706-5668]{R. Chornock}
\UCB
\author[0000-0003-1792-2338]{T. Laskar}
\UTAH
\author[0000-0003-4768-7586]{R. Margutti}
\UCB 
\UCBphys
\author[0000-0001-6747-8509]{Y. Yao}
\UCB
\Miller
\author[0000-0001-7007-6295]{Y. Cendes}
\OR
\CfA
\author[0000-0001-6395-6702
]{S. Gomez}
\STSCI
\author[0000-0003-2349-101X]{A. Hajela}
\DARK
\author[0000-0003-1386-7861]{D. R. Pasham}
\MIT


\begin{abstract}
The tidal disruption event (TDE) AT2022cmc represents the fourth known example of a relativistic jet produced by the tidal disruption of a stray star providing a unique probe of the formation and evolution of relativistic jets in otherwise dormant supermassive black holes (SMBHs). Here we present deep, late-time {\it Chandra} observations of AT2022cmc extending to $t_{\rm obs} \approx 400$ days after disruption. Our observations reveal a sudden decrease in the X-ray brightness by a factor of $\gtrsim 14$ over a factor of $\approx 2.3$ in time, and a deviation from the earlier power-law decline with a steepening $\alpha \gtrsim 3.2$ ($F_X \propto t^{-\alpha}$), steeper than expected for a jet break, and pointing to the cessation of jet activity at $t_{\rm obs} \approx 215$ days. Such a transition has been observed in two previous TDEs (\textit{Swift} J1644+57 and \textit{Swift} J2058+05). From the X-ray luminosity and the timescale of jet shutoff, we parameterize the mass of the SMBH in terms of  unknown jet efficiency and accreted mass fraction parameters. Motivated by the disk-jet connection in AGN, we favor black hole masses $\lesssim 10^5 \ \rm M_{\odot}$ (where the jet and disk luminosities are comparable), and disfavor larger black holes (in which extremely powerful jets are required to outshine their accretion disks). We additionally estimate a total accreted mass of $\approx 0.1 \rm \  M_{\odot}$. Applying the same formalism to \textit{Swift} J1644+57 and \textit{Swift} J2058+05, we favor comparable black hole masses for these TDEs of $\lesssim$ a few $\times 10^5 \ \rm M_{\odot}$, and suggest that jetted TDEs may preferentially form from lower mass black holes when compared to non-relativistic events, owing to generally lower jet and higher disk efficiencies at higher black hole masses. 
\end{abstract}

\keywords{Tidal disruption events; relativistic jets; accretion disks; transients}

\section{Introduction}
\label{sec:intro}

The tidal disruption of a stray star by a supermassive black hole (SMBH) offers a unique opportunity to study the full life-cycle of jets and outflows powered by black holes \citep{Giannios2011,DeColle2020,Dai2021}. Indeed, a small fraction of tidal disruption events (TDEs) discovered to date have been found to harbor powerful relativistic jets \citep{Bloom2011,Burrows2011,Levan2011,Zauderer2011,Cenko2012,Brown2015,Andreoni2022, Pasham2023}. The precise mechanisms governing the production and evolution of jets in TDEs are poorly understood, although prevailing theories invoke jet launching via the Blanford-Znajek (BZ) mechanism \citep{Blandford1977} in which spin energy is extracted from a rapidly spinning black hole through large-scale magnetic fields \citep{Tchekhovskoy2014,Kelley2014}. However, the reservoir of magnetic flux available for producing such strong magnetic fields in otherwise quiescent systems (e.g., from the disrupted star or a pre-existing fossil accretion disk) remains an open question (e.g., \citealt{Kelley2014,Guillochon2017,Bonnerot2017}).

The discovery of two $\gamma$-ray transients by the \textit{Swift}/Burst Alert Telescope (BAT) in 2011 provided the first unambiguous cases of relativistic jets in TDEs \citep{Levan2011,Burrows2011,Bloom2011,Zauderer2011,Cenko2012}. Following their discoveries, \textit{Swift} J164449.3+573451 (hereafter Sw J1644+57) and \textit{Swift} J2058.4+0516 (hereafter Sw J2058+05) were localized to the nuclei of distant galaxies at $z=0.354$ and $z = 1.1853$ respectively, and were notably similar in their early time behavior in the radio, infrared, and X-ray bands. In both cases, the early X-ray emission exhibited rapid variability on timescales as short as $\lesssim 500$ seconds \citep{Brown2015,Mangano2016}, and X-ray luminosities several orders of magnitude above the Eddington limit for a $\sim 10^6 \ \rm M_{\odot}$ black hole, below the maximum allowed mass of $\sim 10^8 \ \rm M_{\odot}$ for TDE flares. For Sw J1644+57, the discovery of bright radio-to-millimeter synchrotron emission independently established the presence of a collimated relativistic jet with a Lorentz factor of $\Gamma \sim$ few \citep{Zauderer2011,Berger2012}. Sw J1644+57's X-ray luminosity subsequently declined following roughly a $t^{-5/3}$ power law decay \citep{Mangano2016}, as expected for fallback accretion in a TDE \citep{Rees1988}, while Sw J2058+05 had a steeper decay with $t^{-2.2}$. A third jetted TDE, Swift J1112.2$-$8238 (hereafter Sw J1112$-$82; $z = 0.89$), was discovered in June 2011 \citep{Brown2015}, although limited data are available for this event. 

A remarkable feature of the first two jetted TDEs is a sudden drop in their X-ray light curves at late times, marking a fundamental change in the nature of the X-ray emission. Deep \textit{XMM-Newton} and {\it Chandra} follow-up observations of Sw J1644+57 showed a precipitous decline in the X-ray luminosity by a factor of $\approx 170$ beginning about $t_{\rm rest} \sim 370$ days after the discovery and over a timescale of only $\sim 70$ days \citep{Zauderer2013}. For Sw J2058+05, the X-ray luminosity declined by a factor of $\sim 160$ at $t_{\rm rest} \sim 250$ days over a similar span of $\lesssim 70$ days \citep{Pasham2015}. In the case of Sw J1112-82, a sharp decline in the X-ray flux was observed at $t_{\rm rest} \sim 20$ days, followed by a non-detection at $\sim 500$ days, although this is not inconsistent with the order of magnitude variability observed in the X-ray light curve for Sw J1644+57 \citep{Saxton2012}.  The sudden decrease in X-ray flux observed for Sw J1644+57 and Sw J2058+05 has been attributed to the cessation of jet activity as the accretion state transitions from super- to sub-Eddington accretion \citep{Zauderer2013,Pasham2015}, providing novel constraints on the properties of the disrupted star and SMBH \citep{Tchekhovskoy2014}.

\begin{deluxetable*}{lccccc}
\tablecolumns{6}
\caption{\textit{Chandra} X-ray Observations of AT2022cmc}
\tablehead{
\colhead{Observation Epoch} &
\colhead{Exposure Time} & 
\colhead{$t^{\rm a}_{\rm obs}$} &
\colhead{$t^{\rm b}_{\rm rest}$} & 
\colhead{Net Count Rate ($0.5 - 8$ keV)} &
\colhead{Flux$^{\rm c}$ ($0.3 - 10$ keV)}\\
\colhead{(MJD)} & 
\colhead{(ks)} & 
\colhead{(days)} & 
\colhead{(days)} & 
\colhead{($\rm cts \ s^{-1}$)} &
\colhead{($\rm erg \ cm^{-2} \ s^{-1}$)}
}  
\startdata
2023 Feb 24 & 30.17 & 378 & 172 & $<2.19\times 10^{-4}$ & $<4.16\times 10^{-15}$\\
2023 Mar 28 & 15.37 & 410 & 187 & $<4.30\times 10^{-4}$ & $<8.16\times 10^{-15}$\\
2023 Mar 29 & 44.58 & 411 & 187 & $<1.48 \times 10^{-4}$ & $<2.81\times 10^{-15}$\\
\hline
2023 Mar $18^{\rm d}$ & 90.12 & 400 & 182 &$<7.33\times10^{-5}$ & $<1.39\times 10^{-15}$ 
\enddata
\tablecomments{Limits correspond to $3\sigma$.\\
$^{\rm a}$ Observer frame.\\
$^{\rm a}$ Rest-frame.\\
$^{\rm c}$ Absorbed flux.\\
$^{\rm d}$ Average exposure-weighted observation epoch corresponding to merged observation.}
\label{tab:chandra}
\end{deluxetable*}

Over a decade since the discoveries of the first jetted TDEs, observations have revealed that relativistic jets in TDEs are extremely rare \citep{Alexander2020}. Compared to the rate of non-jetted TDEs ($\sim 10^3\ \rm Gpc^{-3} \ yr^{-1}$; \citealt{Stone2020}), TDEs that appear to power on-axis relativistic jets comprise less than $1\%$ of the TDE population \citep{Sun2015,Andreoni2022}, suggesting that such events may require special conditions (e.g., high black hole spins or a strong magnetic flux threading the black hole; \citealt{Tchekhovskoy2014}) or implicate viewing angle effects in which the majority of jets are beamed out of our line of sight. Indeed, the emergence of delayed radio emission from a subset of TDEs (e.g., \citealt{Horesh2021b,Cendes2023}) can potentially be explained by off-axis jets \citep{Sfaradi2024,Sato2024}, though non-relativistic outflows provide a viable mechanism as well. Nonetheless, the fraction of powerful jets similar to Sw J1644+57 still appears to be extremely small. The apparent diversity in the X-ray spectra of jetted and non-jetted TDEs \citep{Auchettl2018}, with the former exhibiting generally harder X-ray emission, may be naturally reconciled by the observer's viewing angle with respect to the accretion disk \citep{Dai2018}. Alternatively, a misalignment between the black hole spin axis and orbital plane of the star may produce a quasi-spherical outflow that prevents all but the most powerful and highly magnetized jets from escaping, leading to the observed dichotomy \citep{Lu2023,Teboul2023}.

The discovery of the relativistic TDE AT2022cmc on 2022 February 11 by the Zwicky Transient Facility (ZTF) marked the first jetted TDE discovered in the optical and the first observed to launch a relativistic jet in the last decade. Unlike the \textit{Swift}-detected events, AT2022cmc was initially discovered as a fast-fading optical transient \citep{Andreoni2022_gcn}. Subsequent spectroscopic observations identified a redshift of $z=1.193$ \citep{Tanvir2022}, heralding AT2022cmc as the furthest TDE discovered to date. The early time X-ray emission from AT2022cmc was highly variable \citep{Pasham2023}, similar to the early X-ray light curve of Sw J1644+57, and was interpreted as synchrotron radiation due to energy dissipation within a magnetically dominated jet \citep{Yao2024}. Observations at radio and submillimeter wavelengths further supported the presence of a relativistic jet expanding into an ambient medium \citep{Andreoni2022,Pasham2023,Rhodes2023}, as with Sw J1644+57. 

As one of a small sample of jetted TDEs, AT2022cmc affords a rare opportunity to gain insight into the emission mechanism operating in the accretion disk at early times \citep{Pasham2023,Yao2024}, as well as the properties of the SMBH and disrupted star through long-term monitoring of the X-ray light curve. Here we present late-time {\it Chandra} X-ray observations of AT2022cmc. Proceeding under the assumption that the X-ray emission from AT2022cmc is dominated by a relativistic jet, our observations demonstrate that the jet has shut off at $t_{\rm rest} \sim 100$ days, as evidenced by a sudden drop in the X-ray luminosity and a deviation from an earlier power-law decline. The X-ray data allow us to uniquely probe the mass of the disrupting SMBH and the total accreted mass. We adopt the discovery epoch MJD 59621.4463 (2022 February 11 at 10:42:40 UTC) as the reference epoch. Throughout the paper, we use the Planck cosmological parameters for a flat $\Lambda$CDM universe, with $H_{0}$ = 67.66 km s$^{-1}$ Mpc$^{-1}$, $\Omega_m = 0.310$, and $\Omega_{\lambda} = 0.690$ \citep{Planck2020}.

\section{Chandra X-ray Observations}

We obtained three epochs of X-ray observations of AT2022cmc with the {\it Chandra} Advanced CCD Imaging Spectrometer (ACIS-S; Obs IDs: 26791, 26792, 27769; PI: Eftekhari)\footnote{This paper employs a list of {\it Chandra} datasets, obtained by the {\it Chandra} X-ray Observatory, contained in~\dataset[DOI: 266]{https://doi.org/10.25574/cdc.266}.} on UT 2023 February 24, March 28, and March 29 with a total exposure time of 90.12 ks. Details of the observations and individual exposure times are given in Table~\ref{tab:chandra}. We analyze the data using the \texttt{CIAO} software package (v4.13), following standard ACIS data filtering. We do not detect X-ray emission at the position of AT2022cmc using \texttt{wavdetect} in any of the individual epochs. We align the individual epochs to a common astrometric solution using \texttt{wcs\_match} and \texttt{wcs\_update} and sources identified in our longest exposure image (Obs ID: 27769) using \texttt{wavdetect} as a reference. We then merge the observations to generate a co-added, exposure-corrected image using \texttt{merge\_obs}. We do not detect any counts in a 1$\arcsec$ radius aperture centered at the position of AT2022cmc in the merged event file, corresponding to a $3\sigma$ upper limit on the $0.5 - 8$ keV count rate of $7.3\times 10^{-5}\ \rm cts \ s^{-1}$ (assuming Poisson statistics; \citealt{Gehrels1986}). Adopting a photon index of $\Gamma=1.6$ (the best fit photon index from fits to \textit{Swift}/XRT data\footnote{There is minimal evolution in the \textit{Swift}/XRT X-ray spectrum at $t_{\rm rest} \sim 2 - 10$ days, with the best fit photon index varying between $\Gamma \sim 1.3 - 1.9$.}; \citealt{Pasham2023}), $\rm N_{H,MW} = 8.8\times 10^{19} \ cm^{-2}$ \citep{Kalberla2005}, and $\rm N_{H,int} =  10^{21} \ cm^{-2}$ \citep{Yao2024}, the $3\sigma$ limit on the observed $0.3-10$ keV absorbed X-ray flux\footnote{https://cxc.harvard.edu/toolkit/pimms.jsp} is $F_X \lesssim  1.39 \times 10^{-15} \ \rm erg \ cm^{-2} \ s^{-1}$. We note that the X-ray spectrum of AT2022cmc is consistent with synchrotron emission from a magnetically dominated jet \citep{Yao2024}, similar to that of Sw J1644+57 \citep{Burrows2011}.

\begin{figure*}
\includegraphics[width=\textwidth]{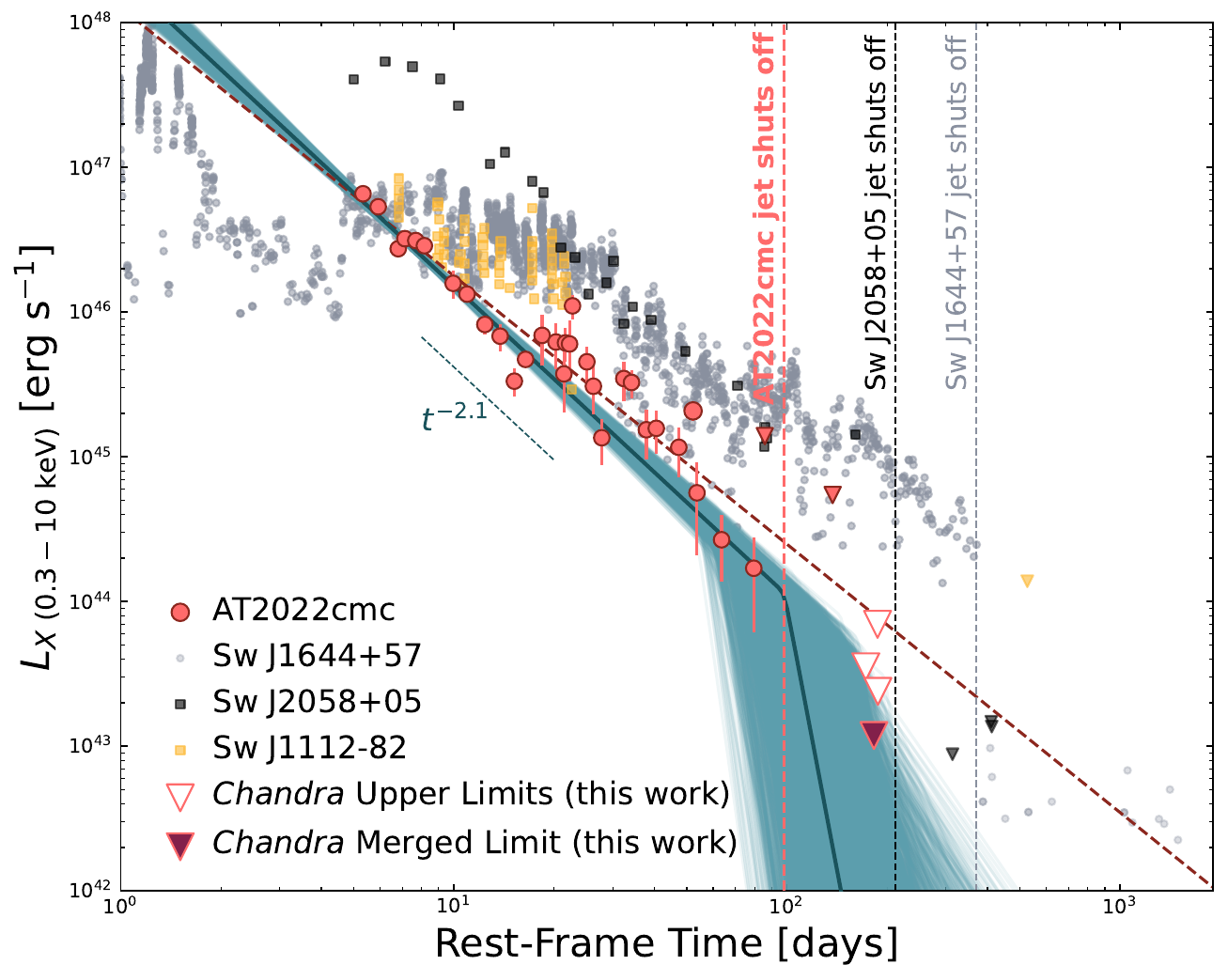}
\caption{Stringent late-time upper limits on the X-ray emission of AT2022cmc obtained in this work place robust constraints on the shut-off time of the relativistic jet. Our late-time {\it Chandra} observations (large open triangles) and our merged {\it Chandra} epoch (large solid triangle) at $t_{\rm rest} \approx 182$ days indicate a deviation from the earlier power-law decline (dashed maroon line) in the X-ray light curve based on \textit{Swift}/XRT ($t_{\rm rest} \sim 5 - 86$ days; \citealt{Yao2024}) and \textit{XMM-Newton} ($t_{\rm rest} = 52$ and $137$ days; \citealt{Yao2024}) data. Our best fit broken power-law model is shown as a dark blue line, while light blue lines represent random samples from the MCMC chains. 
Also shown for comparison are the X-ray light curves of other jetted TDEs, including Sw J1644+57 (light gray circles; \citealt{Burrows2011,Zauderer2013,Eftekhari2018}), Sw J2058+05 (dark gray squares; \citealt{Pasham2015}), and Sw J1112-82 (yellow squares; \citealt{Brown2015}). Vertical dashed lines indicate the time of jet shut-off for each event.}
\label{fig:xraylc}
\end{figure*}

We plot our X-ray limits for AT2022cmc, including our merged limit (with an average exposure-weighted observation epoch of $t_{\rm rest} \approx 182$ days post-disurption) and limits for individual epochs in Figure~\ref{fig:xraylc}. We also plot X-ray data collected by \textit{Swift}/XRT between MJD 59633 and MJD 59810 ($t_{\rm rest} \sim 5 - 86$ days) as compiled in \citealt{Yao2024} and data from \textit{XMM-Newton} (with a best fit photon index $\Gamma = 1.65$; \citealt{Yao2024}), including a detection and upper limit at $t_{\rm rest} \sim 52$ and $137$ days, respectively. We note that based on experience, the \textit{Chandra}/ACIS-S vs. \textit{Swift}/XRT inter-calibration should impart negligible variations to the light curve of order $<$15\%.

\section{Results}
\subsection{The Relativistic Jet Shuts Off}
\label{sec:model}

The X-ray light curve of AT2022cmc (Figure~\ref{fig:xraylc}) exhibits rapid ($\sim 10^3\ \rm s$) variability on timescales of $\sim$weeks after its initial discovery \citep{Pasham2023} and an overall $F_X \propto t^{-1.9}$ decline to $t_{\rm rest} \approx 80$ days. The $t^{-1.9}$ decline is somewhat steeper than the canonical $t^{-5/3}$ decay for TDEs and may be indicative of a partial stellar disruption in which the stellar core survives and is able to partially oppose the black hole's gravitational force \citep{Guillochon2013}. In particular, the deviation from $t^{-5/3}$ on the observed timescale of $t_{\rm rest} \lesssim 100$ days may imply a substantial stellar core in which $\gtrsim 15\%$ of the initial stellar mass is left behind \citep{Coughlin2019}. Our deep \textit{Chandra} limit at $t_{\rm rest}\sim 182$ days indicates a decrease in the X-ray flux by a factor of $\gtrsim 14$ and a deviation from the earlier power-law decline with a steepening $\alpha \gtrsim 3.2$\ ($F_X \propto t^{-\alpha}$) pointing to a change in the nature of the X-ray emission,\footnote{We note that the steepening in the X-ray light curve is inconsistent with a jet break for which we expect a shallower power-law index \citep{Sari1999,Wang2018}.} which we attribute to the cessation of jet activity. 

To quantify this, we fit a smoothed broken power-law of the form
\begin{equation}
F_X(t) = F_X \bigg[\bigg(\dfrac{t}{t_{\rm off}}\bigg)^{-s \alpha_1} + \bigg(\dfrac{t}{t_{\rm off}}\bigg)^{-s \alpha_2} \bigg]^{-1/s}
\end{equation} 
to the X-ray light curve using \texttt{emcee} \citep{Foreman-Mackey2013}, a Python-based implementation of a Markov Chain Monte Carlo (MCMC) Ensemble Sampler. We fix the smoothing parameter $s=10$ and fit for the break point $t_{\rm off}$, the flux normalization $F_X$ at $t_{\rm off}$, and the power-law indices before ($\alpha_1$) and after ($\alpha_2$) the break, and include an additional parameter which accounts for an underestimate of the variance in the light curve by a constant factor $f$. 

To incorporate the upper limits into the fits, we adopt the prescription from \citet{Laskar2014} in which the likelihood function accounts for both detections and non-detections in a given data set and is given by \citep{Lawless2002,Helsel2005}:
\begin{equation}
\mathcal{L} = \prod p(e_i)^{\delta_i} F(e_i)^{(1-\delta_i)}
\end{equation}
where $e_i$ are the residuals between the predicted model flux and the measurement or $3\sigma$ upper limit, $p(e_i)$ and $F(e_i)$ are the probability density and cumulative distribution functions of the residuals, respectively, and $\delta_i$ indicates an upper limit ($\delta_i = 0$) or a detection ($\delta_i = 1$). We approximate the measurement uncertainties ($\sigma_i$) as Gaussian by taking the mean of the asymmetric errors on each data point and adopt for the non-detections the Poisson single-sided upper limits. The probability density and cumulative distribution functions are therefore given by
\begin{equation}
p(e_i) = \dfrac{1}{\sqrt{2\pi}\sigma} \exp^{-e_i^2/2\sigma_i^2}
\end{equation}
and
\begin{equation}
F(e_i) = \dfrac{1}{2}\bigg[1+{\rm erf}\bigg(\dfrac{e_i}{\sqrt{2}\sigma_i}\bigg)\bigg],
\end{equation}
respectively, where erf is the error function.

\begin{figure}
\includegraphics[width=\columnwidth]{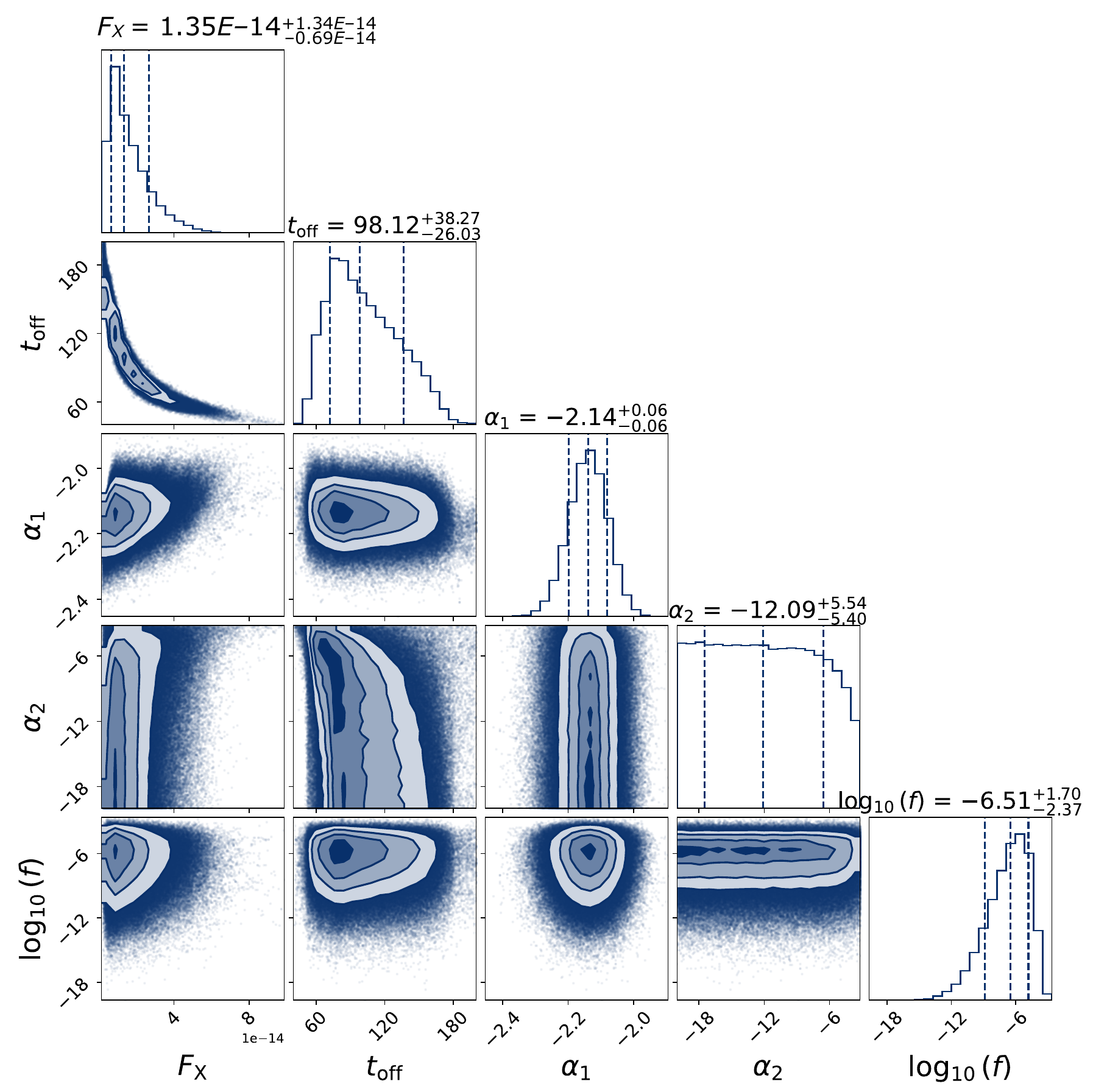}
\caption{Results from our MCMC parameter estimation for a broken power law fit to the X-ray light curve of AT2022cmc. Marginalized posterior distributions for each parameter are shown on the diagonal, where dashed lines indicate the median and $68\%$ confidence interval.}
\label{fig:corner}
\end{figure}

For the priors, we use log-uniform priors for $F_X$ and $t_{\rm off}$, fixing the allowed range for $t_{\rm off}$ to span the full timescale of the X-ray light curve ($t_{\rm rest }\sim 0 - 200$ days) and allowing for a possible break at earlier times as evidenced by an apparent steepening at $t_{\rm rest} \approx 50$ d (see Figure~\ref{fig:xraylc}). We use linearly uniform priors for both power-law indices with $-5 < \alpha_1 < 0$ and $-15 < \alpha_2 < -3.2$, where the upper bound on $\alpha_2$ is set by the slope between the last \textit{Swift}/XRT detection and our deep \textit{Chandra} upper limit. The posterior distributions are sampled using 1000 Markov chains and 95000 steps, where we discard the first 10000 steps ($\approx 10\times$ the integrated autocorrelation length) to ensure the walkers have sufficiently converged and that the samples represent independent, uncorrelated measures of the target distribution. We further assess the quality of our MCMC samples by calculating the acceptance fraction of the ensemble and find a mean acceptance fraction of 0.22, within the nominally accepted range \citep{Gelman1996}. The best fit parameters and posterior distributions are given in Table~\ref{tab:mcmc} and Figure~\ref{fig:corner}, respectively.

Based on the results of our MCMC analysis, we find evidence for a break in the X-ray light curve at $t_{\rm off} \equiv t_{\rm rest} = 98^{+38}_{-26}$ days ($t_{\rm obs} \approx 215$ days). We note a degeneracy between the jet shut-off time $t_{\rm off}$ and the flux normalization $F_X$ given the sparse temporal coverage in the light curve between the last detection and our late-time \textit{Chandra} limits. Compared to Sw J2058+05 and Sw J1644+57, the two known jetted TDEs with observed X-ray drop-offs at $t_{\rm rest} \approx 200$ and $370$ days, respectively, the drop in AT2022cmc's X-ray light curve occurs a factor of two to four times earlier. We explore the implications of this in Section~\ref{sec:mbh_time}.

\subsection{Estimating the Mass of the Disrupting SMBH: X-ray Luminosity}
\label{sec:mass}

The abrupt decrease in the X-ray luminosity at $t_{\rm rest} \approx 100$ days for AT2022cmc can be interpreted as an accretion state transition from super- to sub-Eddington, providing a novel estimate for the mass of the disrupting SMBH. Such a transition is predicted from numerical simulations as the accretion disk becomes geometrically thin and radiatively efficient \citep{DeColle2012}. Here we leverage the observational data for AT2022cmc to estimate the mass of the disrupting SMBH.

To first order, we can estimate $M_{\rm BH}$ by equating the X-ray luminosity at turnoff to the Eddington luminosity. Assuming the jet shuts off at an Eddington ratio $\lambda=\dot{M}/\dot{M}_{\rm Edd} = 1$ ($\dot{M}_{\rm Edd} = L_{\rm Edd}/\epsilon_{\rm disk} c^2$), the disk luminosity at shut-off ($L_{\rm disk, off} = \epsilon_{\rm disk} \dot{M}_{\rm BH} c^2$) is equal to the Eddington luminosity, and we can therefore parameterize the black hole mass as
\begin{equation}
M_{\rm BH} = 10^5 \dfrac{L_{\rm jet,off}}{1.19\times10^{45} \ \rm erg \ s^{-1}} \bigg(\dfrac{\epsilon_{\rm disk}}{\epsilon_{\rm jet}}\bigg)\ f_{\rm beam,200}^{-1} f_{\rm bol,3}
\rm\ M_{\odot}
\label{eq:mbh}
\end{equation}
where $L_{\rm jet,off}  = \epsilon_{\rm jet} \dot{M}_{\rm BH} c^2$ is the isotropic X-ray luminosity at jet shut-off and is equal to $L_{\rm disk,off} (\epsilon_{\rm jet}/\epsilon_{\rm disk})$ for $\lambda=1$, and  $\epsilon_{\rm jet}$ and $\epsilon_{\rm disk}$ are the jet and disk radiative efficiencies, respectively. We convert the observed isotropic X-ray luminosity $L_{\rm jet,off}$ into an intrinsic jet luminosity $L_{\rm jet}$ for a ``standard'' jet assuming a relativistic beaming correction $f_{\rm beam} = 200f_{\rm beam,200}$ for a beaming angle of 0.1 rad (as estimated for Sw J1644+57) and a bolometric correction $f_{\rm bol} = 3f_{\rm bol,3}$ (as in previous work for jetted TDEs; \citealt{Burrows2011,Pasham2015}) to increase the observed X-ray luminosity by a factor of $3$.

\begin{deluxetable}{lc}
\tablecolumns{2}
\caption{AT2022cmc X-ray Light curve Broken Power Law Fit}
\tablehead{
\colhead{Parameter} &
\colhead{Best Fit} 
}  
\startdata 
$F_{X\rm \ (0.3-10 \ keV)}$ ($\rm erg \ cm^{-2} \ s^{-1}$) & $1.35^{+1.3}_{-0.7}\times 10^{-14}$\\
$t_{\rm off}^{\rm a}$ (d) & $98^{+38}_{-26}$\\
$\alpha_1$ & $-2.1^{+0.1}_{-0.1}$\\
$\alpha_2$ & $-12.1^{+5.5}_{-5.4}$\\
\enddata
\tablecomments{$^{\rm a}$Rest-frame time.}
\label{tab:mcmc}
\end{deluxetable}

\begin{figure*}
\includegraphics[width=\textwidth]{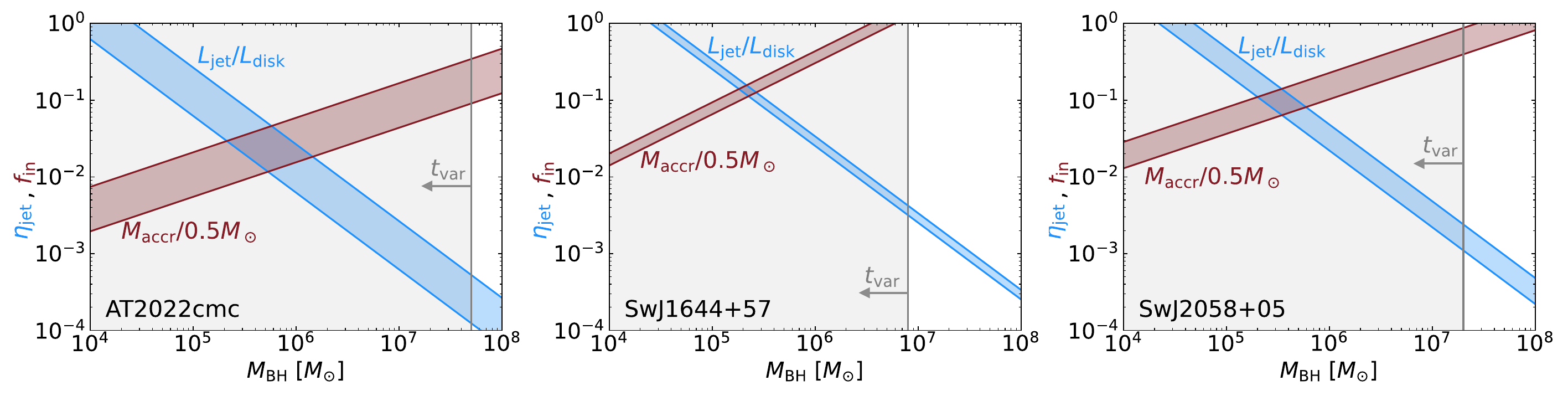}
\caption{Allowed phase space of black holes masses as a function of both the jet efficiency ($\eta_{\rm jet} = \epsilon_{\rm jet}/\epsilon_{\rm disk}$) and the fraction of accreted matter ($f_{\rm in} = f_{\rm accr} f_{\rm circ} f_{\rm partial}$) for AT2022cmc, SwJ1644+57, and SwJ2058+05 where individual curves correspond to the ratio of intrinsic jet and disk luminosity ($L_{\rm jet}/L_{\rm disk}$ with $L_{\rm jet} = L_{\rm jet,off} f_{\rm beam,200}^{-1} f_{\rm bol,3}$) at jet shut-off (blue curves) and the total accreted mass relative to a complete disruption ($M_{\rm accr}/0.5 M_{\odot}$; red curves). The width of each curve corresponds to the uncertainty in the X-ray luminosity and timescale of jet shut-off. For a given black hole mass, there exists a one-to-one mapping between $f_{\rm in}$ and $\eta_{\rm jet}$. The gray shaded regions denotes the upper limit constraints on $M_{\rm BH}$ for each source from X-ray variability arguments.}
\label{fig:m_bhs}
\end{figure*}

In Figure~\ref{fig:m_bhs}, we plot $\eta_{\rm jet} \equiv \epsilon_{\rm jet}/\epsilon_{\rm disk}$ as a function of  $M_{\rm BH}$ using equation~\ref{eq:mbh} where we set $L_{\rm jet,off}$ equal to the break luminosity at the time of jet shut-off from our MCMC fit ($L_X \approx 10^{44}\, \rm erg \ s^{-1}$). The result is a curve in the $\eta_{\rm jet}-M_{\rm BH}$ parameter space, corresponding to the ratio of intrinsic\footnote{Here we consider the intrinsic jet luminosity as an approximation for a ``standard'' jet with $\theta_j \sim 0.1$ rad given the unknown jet opening angle.} jet and disk luminosities at jet shut-off ($\eta_{\rm jet} \equiv L_{\rm jet,off} f_{\rm beam,200}^{-1} f_{\rm bol,3}/L_{\rm disk,off}$). We repeat the above exercise for Sw J1644+57 and Sw J2058+05, the two other jetted TDEs with observed X-ray drop-offs, and plot the results in Figure~\ref{fig:m_bhs}. For Sw J2058+05, we re-fit the X-ray light curve to better constrain $t_{\rm off}$ (see Appendix~A). In each case, the width of the curve corresponds to the uncertainty in the X-ray luminosity at jet shut-off.

Studies of the disk-jet connection in AGN, including the most powerful blazars, indicate that the power in relativistic jets is strongly coupled to their accretion disk luminosities \citep{Ghisellini2010,Inoue2017}. The intrinsic power radiated by the jet in the form of non-thermal luminosity (a factor of at least 3--10 times smaller than the jet's bulk kinetic power) 
is quite large, and in some cases equal to, or only a factor of a few times smaller than the disk luminosity \citep{Ghisellini2010}. Moreover, theoretical work investigating accretion dynamics and jet power in relativistic TDEs has shown that the jet power exceeds the disk luminosity for reasonable system parameters\footnote{This analysis assumes a low radiative disk efficiency ($\epsilon_{\rm disk} = 0.057$) while at jet shut-off the disk has become radiatively efficient with $\epsilon_{\rm disk} = 0.1$.}  \citep{Piran2015}, and particularly at lower black hole masses \citep{Krolik2012}. Thus motivated, we favor values of $\eta_{\rm jet} \gtrsim 10^{-1}$ for AT2022cmc where the disk and jet luminosities are comparable, and hence black hole masses $\lesssim 10^5 \ \rm M_{\odot}$. We note that our inference of a high jet efficiency is furthermore consistent with the high jet efficiency inferred for AT2022cmc in the unified TDE framework of \citet{Teboul2023} based on the early X-ray peak which implies a promptly escaping jet and rapid magneto-spin alignment with the SMBH spin. We find similar results for both Sw J1644+57 and Sw J2058+05, which exhibit comparable X-ray luminosities at jet shut-off, where $M_{\rm BH}\lesssim$ a few $\times 10^5 \ \rm M_{\odot}$ for $\eta_{\rm jet} \gtrsim 10^{-1}$. We note that larger black hole masses ($\sim 10^{7} \ \rm M_{\odot}$) are disfavored, as these would imply that the jet power is suppressed well below the luminosity of the accretion flow, which we consider contrived, particularly in light of the on-axis orientation for the sample of jetted TDEs. Indeed, even low-luminosity AGN powering relatively weaker jets exhibit a positive disk-jet coupling correlation, albeit shallower than observed for powerful radio galaxies and quasars \citep{Nagar2005,Su2016}.

Our results suggest that jetted TDEs may preferentially form from lower mass black holes when compared to non-relativistic events which comprise the bulk of the TDE population and typically occur in galaxies hosting more massive black holes \citep{Mockler2019,Ryu2020,Nicholl2022,Hammerstein2023a}, similar to the findings of \citealt{Somolwar2023} based on a sample of radio-selected TDEs. If the latter form in systems with larger black hole masses, the paucity of jets may be a direct consequence of the lower jet efficiency and higher disk efficiency in this mass regime. Indeed, a positive correlation between black hole mass and disk radiative efficiency has been suggested in a sample of optically bright TDEs \citep{Nicholl2022}. At high black hole masses, such systems would therefore require extremely powerful jets (i.e., large jet efficiencies) to outshine their accretion disks.

\subsection{Estimating the Mass of the Disrupting SMBH: Timescale of Jet Shut-off}
\label{sec:mbh_time}

Separate from the luminosity at jet shut-off, we can independently infer the mass of the disrupting SMBH from the {\it timescale} of jet shut-off. For a main-sequence star, the fallback accretion rate $\dot{M}$ onto the BH peaks at a timescale $t \approx t_{\rm fb}$ given by (e.g., \citealt{Ulmer1999}):
\begin{equation}
t_{\rm fb} = 0.11\ {\rm yr}\, \bigg(\dfrac{M_{\rm BH}}{10^6\ M_{\odot}}\bigg)^{1/2} \bigg(\dfrac{M_*}{M_{\odot}}\bigg)^{-1}
\label{eq:tfb}
\end{equation}
and subsequently decays following a power law 
\begin{equation}
\dot{M} = \dot{M}_{\rm peak}(t/t_{\rm fb})^{-\alpha}
\end{equation}
where $\alpha=5/3$ for a complete disruption and $\alpha=2.2$ for a partial disruption in which the stellar core survives and $\lesssim 50\%$ of the stellar mass is accreted onto the black hole \citep{Guillochon2013}. The Eddington ratio, $\lambda = \dot{M}/\dot{M}_{\rm Edd}$, therefore evolves as 
\begin{equation}
\lambda = \lambda_{\rm peak, fb}(t/t_{\rm fb})^{-\alpha}.
\label{eq:lambda_vs_t}
\end{equation}
Assuming $100\%$ efficiency in fallback mass reaching the black hole, the peak Eddington ratio is given by \citep{Stone2013}:
\begin{equation}
\lambda_{\rm peak, fb} \simeq 133 \ \epsilon_{\rm disk,-1}\bigg(\dfrac{\alpha -1}{2/3}\bigg)\bigg(\dfrac{M_{\rm BH}}{10^6 \ M_{\odot}} \bigg)^{-3/2} \bigg(\dfrac{M_*}{M_\odot}\bigg)^{2}.
\label{eq:lambdapeakfb}
\end{equation}

In reality, the accretion of matter onto the black hole is not a strictly efficient process \citep{Metzger2016} as some fraction of the initially bound stellar debris becomes unbound due to shocks formed at the self-intersection point of the debris stream \citep{Lu2020}. Moreover, the low gravitational binding energy of the debris, coupled with the fact that the infalling gas cannot radiatively cool when the fallback rate is super-Eddington \citep{Strubbe2009}, implies that only a small fraction of the disrupted stellar material is accreted by the black hole on a viscous timescale \citep{Shen2014,Metzger2016}. We therefore parameterize the disk efficiency in converting the fallback accretion rate, given by eq.~\ref{eq:lambda_vs_t} and \ref{eq:lambdapeakfb}, into luminosity in terms of the fraction of accreted matter $f_{\rm in}= f_{\rm accr} f_{\rm circ} f_{\rm partial}$, where $f_{\rm accr}$ is the accretion efficiency, $f_{\rm circ}$ is the circularization efficiency, and $f_{\rm partial}$ accounts for a partial disruption in which only a fraction of stellar mass is stripped from the star. The Eddington ratio at the black hole is therefore given by:
\begin{equation}
\lambda_{\rm peak, BH} = f_{\rm in} \lambda_{\rm peak, fb}
\label{eq:lambdapeakbh}
\end{equation}
Taking $\lambda = 1$ to mark the transition from super- to sub-Eddington accretion and hence the point at which the jet shuts off, the jet shut-off time is  given by:\begin{equation}
t_{\rm off} = t_{\rm fb}\lambda_{\rm peak,BH}^{1/\alpha}.
\label{eq:toff}
\end{equation}

Using equations \ref{eq:tfb}, \ref{eq:lambdapeakfb}, \ref{eq:lambdapeakbh}, and \ref{eq:toff} with $\alpha = 2.1$ as constrained from our MCMC analysis, our inferred jet shut-off time of $t_{\rm rest} = 98$ days, and assuming a solar mass star, in Figure~\ref{fig:m_bhs} we plot black hole mass as a function of $f_{\rm in}$, where we adopt $\epsilon_{\rm disk} = 0.1$ as a typical radiative efficiency.\footnote{In reality, the radiative efficiency for super-Eddington accretion disks decreases as the Eddington ratio $\lambda$ increases \citep{Jiang2019}. For our purposes, and given that we carry out our calculations at the time of jet shut-off when the radiative efficiency is expected to be close to the typical value, we adopt a constant radiative efficiency $\epsilon_{\rm disk} = 0.1$.} The result is a curve in the $f_{\rm in} - M_{\rm BH}$ phase-space corresponding to the fraction of accreted mass relative to a full disruption, where $f_{\rm in} = 1$ implies a complete disruption in which 0.5 $M_{\odot}$ is accreted onto the black hole.\footnote{Here we assume the tidal disruption of a $1\ M_{\odot}$ main sequence star in which half of the stellar material is unbound from the system as the most plausible scenario and additionally allow for the possibility that the star is only partially disrupted. In the case of a white dwarf disruption, which is tenable for a $10^5 \ \rm M_{\odot}$ black hole and indeed was suggested for Sw J1644+57 based on the short timescale variability in the {\it Swift}/XRT lightcurve \citep{Krolik2011}, in which the radius and mass of the star are inversely related ($r_* \propto M_*^{-1/3}$), the mass accretion rate on the same timescale is approximately $10^3$ times smaller \citep{Tchekhovskoy2014}, implying a larger accreted mass fraction for a fixed black hole mass.}

Given $L_{\rm jet,off}$ and $t_{\rm off}$, for fixed values of $M_{\rm BH}$ in Figure~\ref{fig:m_bhs}, there exist corresponding values of $f_{\rm in}$ and $\eta_{\rm jet}$. Under the above supposition that the jet and disk luminosities are comparable (i.e.  $\eta_{\rm jet} \gtrsim 10^{-1}$), as in powerful blazars and radio-loud quasars, and $M_{\rm BH} \lesssim 10^5 \ \rm M_{\odot}$, $f_{\rm in}$ is constrained to $\lesssim 10^{-2}$ for AT2022cmc, implying a small fraction of total accreted mass relative to a complete disruption in which $0.5 \ \rm M_{\odot}$ is accreted. This is consistent with the framework presented by \citealt{Metzger2016} in which they postulate that $f_{\rm in} \ll 1$ is required to produce the observed outflows in TDEs. On the other hand, if the jet-disk coupling in TDEs is intrinsically distinct from that of AGN, the jet may be radiatively inefficient ($\eta_{\rm jet} \lesssim 10^{-3}$) at high black hole masses ($M_{\rm BH} \sim 10^7 \ \rm M_{\odot}$) despite a large fraction of accreted mass ($f_{\rm in} \gtrsim 0.1$). In other words, the X-ray luminosity would represent a small fraction of the total energy budget of the system. Such a scenario may arise due to differences in the available magnetic reservoir for TDEs \citep{Kelley2014}, for example, or if a large fraction of jets are choked \citep{Lu2023,Teboul2023} or launched off-axis \citep{Dai2018}. Indeed, the latter scenario may be reflected by the smaller fraction of relativistic jetted TDEs relative to the fraction of radio-quiet quasars \citep{Alexander2020}.


Adopting the same methodology for Sw J1644+57\footnote{We adopt $\alpha=5/3$ for Sw J1644+57, consistent with its observed X-ray light curve \citep{Burrows2011}.} and Sw J2058+05\footnote{We constrain the timescale of jet shut-off and the power-law index for Sw J2058+05 in Appendix A.}, we trace out similiar regions in the $f_{\rm in} - M_{\rm BH}$ phase-space in Figure~\ref{fig:m_bhs}. We note that in both cases, the timescale of jet shut-off occurs a factor of two to four times later than in AT2022cmc; requiring the same condition of $\eta_{\rm jet} \gtrsim 10^{-1}$ therefore allows for somewhat larger black hole masses relative to AT2022cmc of $\sim$ a few $\times 10^5 \ \rm M_{\odot}$ and hence a larger fraction of accreted mass $f_{\rm in} \lesssim 0.1$, implying that the total accreted mass in these systems is more comparable to that of a complete disruption.


\subsection{Comparison to Other SMBH Mass Estimates}
\label{sec:mass_comparisons}

We compare our SMBH mass estimates for AT2022cmc and other jetted TDEs as derived from the jet shut-off time and X-ray luminosity to estimates derived via other methods. First, we use the observed variability in the X-ray light curves at early times to place an upper limit on the mass of the SMBH. In particular, we can equate the size of the X-ray emitting region $r_s$ to the Schwarzschild radius of a black hole with mass $M_{\rm BH}$. For an observed variability timescale $t_{\rm var,sec} \sim r_s/c$ in seconds, the mass of the SMBH is given by
\begin{equation}
    M_{\rm BH} \lesssim 10^5 \ M_{\odot} \dfrac{t_{\rm var, sec}}{(1+z)}
\end{equation}

In the case of AT2022cmc, X-ray variability is observed over a wide range of timescales spanning $1000$ s to $\sim$ days. We adopt a minimum variability timescale $t_{\rm var} = 1000$ s \citep{Pasham2023}. 
The resulting limit on the SMBH mass for AT2022cmc is $M_{\rm BH} \lesssim 5\times10^{7} \ M_{\odot}$. 

For Sw J1644+57, the early time X-ray light curve out to $t_{\rm obs}\approx 10$ days is punctuated by rapid variability on timescales as short as $\sim 100$ s \citep{Mangano2016}. We therefore constrain the black hole mass for Sw J1644+57 based on variability arguments to $M_{BH} \lesssim 8\times10^{6} \ \rm M_{\odot}$, as in \citet{Burrows2011}. Conversely, the putative detection of quasi-periodic oscillations (QPOs) in the X-ray light curve of Sw J1644+57 \citep{Reis2012} enabled constraints on the black hole mass of $10^5 - 10^6 \ \rm M_{\odot}$ \citep{Abramowicz2012}, comparable to the black holes mass we infer in our jet shut-off paradigm. For Sw J1112$-$82, the X-ray light curve exhibited variability on timescales of a few thousand seconds \citep{Brown2015}, corresponding to a limit on the SMBH mass of $\lesssim 2\times 10^7 \ \rm M_{\odot}$. Finally, for Sw J2058+05, the observed variability timescale of $\sim 500$ s constrains the SMBH mass to $\lesssim 2\times 10^8 \ \rm M_{\odot}$. This is above the maximum allowed mass for TDEs of $<10^8 \ \rm M_{\odot}$, above which the disruption radius lies within the Schwarzschild radius and the star is thus swallowed whole \citep{Rees1988}, and hence is not constraining.


Finally, we note that several limits have been placed on the SMBH mass for jetted TDEs using the galaxy bulge -- black hole mass relation \citep{McConnell2013}. In the case of AT2022cmc, the host galaxy is not detected in deep ground-based imaging with the Canada-French-Hawaii Telescope down to a limiting $r$-band magnitude of 24.5 ($3\sigma$), corresponding to an upper limit on the host galaxy mass of $<1.6\times 10^{11} \ \rm M_{\odot}$ and hence an SMBH mass of $M_{\rm BH} < 4.7 \times 10^8 \ \rm M_{\odot}$ \citep{Andreoni2022}. Similarly derived constraints for Sw J1644+57 and Sw J2058+05 have led to SMBH mass limits of $M_{\rm BH}< 2\times 10^7 \ \rm M_{\odot}$ \citep{Burrows2011} and $M_{\rm BH} < 3 \times 10^7 \ \rm M_{\odot}$ \citep{Pasham2015}, respectively. For Sw J1112-82, \citet{Brown2015} obtain an SMBH mass limit of $2 \times 10^6 \ \rm M_{\odot}$ using the black hole mass - bulge mass relation of \citet{Haring2004}.

In general, we find that our SMBH mass estimates as inferred from the timescale and luminosity of jet shut-off are consistent with estimates derived via other methods, but that such methods do not sufficiently probe the low mass regime where our models are favored.

\subsection{Constraining the Mass Accretion Rate and the Total Accreted Mass}
\label{sec:accretion}

While the precise nature of relativistic jet production in TDEs is not well understood, such jets are canonically expected to form when the accretion rate onto the black hole exceeds the Eddington rate \citep{Giannios2011}. 
The low occurrence rate of jetted TDEs however suggests that super-Eddington accretion, which is expected at early times for disruptions of solar mass stars by black holes with $M_{\rm BH} \lesssim 10^8 \ \rm M_{\odot}$ \citep{DeColle2012}, is not a sufficient condition for powering relativistic jets, and that additional parameters such as black hole spin or disk-jet alignment must play a role \citep{Stone2012,Tchekhovskoy2014,Franchini2016,Curd2019,Zanazzi2019,Teboul2023}. Nevertheless, the rapid drop in flux observed in the X-ray light curves of jetted TDEs on timescales commensurate with a transition to sub-Eddington accretion as seen in numerical simulations supports a connection between super-Eddington accretion and jet production \citep{DeColle2012}.

With the inference that the accretion rate at $t_{\rm rest} =  100$ days is equal to the Eddington accretion rate, we can estimate the total accreted mass onto the SMBH for AT2022cmc. First, we calculate the Eddington accretion rate assuming $L_{\rm Edd}$ is equal to the isotropic X-ray luminosity at the time of jet shut-off ($L_X \approx 10^{44}\, \rm erg \ s^{-1}$) and adopting a beaming angle of 0.1 rad and a bolometric correction factor of 3 as before. We find $\dot{M}({\rm 100 \ days}) \approx \dot{M}_{\rm Edd} \approx 0.003 \ \rm M_{\odot} \ yr^{-1}$ for a radiative efficiency $\epsilon_{\rm disk}=0.1$. Following the prescription of \citet{Zauderer2013}, a simple model for the mass accretion rate is given by $\dot{M}(t) =\dot{M}_p (t/t_j)^{-\alpha}$, where $\dot{M}_p$ is the peak accretion rate and $t_j$ is the timescale over which the X-ray luminosity is constant. We take $t_j = 2.5$ days given the onset of observable X-ray emission on this timescale and the lack of a plateau in the X-ray light curve \citep{Andreoni2022,Pasham2023} and $\alpha=2.1$ and find $\dot{M}_p \approx 4.6 \ \rm M_{\odot} \ yr^{-1}$. Integrating the mass accretion rate out to $t_{\rm off} = 100$ days where $\dot{M}(t) = \dot{M}_p$ at $t<2.5$ days and $\dot{M}(t) =\dot{M}_p (t/t_j)^{-2.1}$ at $t \gtrsim 2.5$ days, we find a total accreted mass of $\approx 0.1 \ \rm M_{\odot}$. 

Our estimate of the total accreted mass is larger than inferred from Figure~\ref{fig:m_bhs}, where $M_{\rm BH} \sim 10^{5} \ \rm M_{\odot}$ implies a total accreted mass $M_{\rm accr} \sim f_{\rm in} \times 0.5 \ \rm M_{\odot} \approx 10^{-2} \times 0.5 \ M_{\odot} \approx 0.01 \ \rm M_{\odot}$. However, as discussed in Section~\ref{sec:mbh_time}, in Figure~\ref{fig:m_bhs}, we account for additional efficiency factors ($f_{\rm in}= f_{\rm accr} f_{\rm circ} f_{\rm partial}$) for converting the fallback accretion into luminosity, which we neglect in our above calculation. Moreover, we consider our above estimate of $0.1 \ \rm M_{\odot}$ approximate, given uncertainties in the assumed radiative efficiency which depends on the spin of the black hole (e.g., \citealt{Novikov1973,Shakura1973,Sadowski2016}). Similarly, if the jet instead shuts off at a fraction of the Eddington accretion rate (e.g., \citealt{Tchekhovskoy2014}), the total accreted mass will decrease linearly with this fraction. We thus consider our two independent estimates of the total accreted mass as broadly consistent with one another.

We repeat the above calculations for Sw J1644+57 and Sw J2058+05 adopting the same beaming and bolometric correction factors of 0.1 rad and 3, respectively. For Sw J1644+57, we take $t_j = 15$ days in the observer frame based on the observed X-ray plateau \citep{Burrows2011} and $\alpha=5/3$ and find a peak accretion rate of $\dot{M}_{\rm peak}\sim 2.4 \ \rm M_{\odot} \ yr^{-1}$. The accretion rate at $t\lesssim 15$ days is therefore given by $\dot{M}(t)= \dot{M}_p$ and $\dot{M}(t) =\dot{M}_p (t/t_j)^{-5/3}$ at $t \gtrsim 15$ days. Integrating the mass accretion rate out to $t_{\rm off}\approx 370$ days (rest-frame), the total accreted mass is $\approx 0.17 \ \rm M_{\odot}$, consistent with the results of \citet{Zauderer2013}. For Sw J2058+05, we adopt $t_j = 11$ days \citep{Cenko2012} and $\alpha=2.1$ and find $\dot{M}_p = 3.5 \ \rm M_{\odot} \ yr^{-1}$. The total mass accreted out to $t_{\rm off} \approx 212$ days (rest-frame) is therefore $\approx 0.2 \ \rm M_{\odot}$. 

\subsection{Revival of the Jet}

As the fallback accretion rate continues to decline following $\dot{M} \propto t^{-2.1}$ and eventually reaches a few percent of Eddington, the disk is expected to transition to a radiatively inefficient advection-dominated accretion flow \citep{Maccarone2003}, analagous to the ``low/hard'' state in X-ray binaries \citep{Fender2004,Russell2011}. At this point, the jet is expected to turn back on. Based on the Eddington accretion rate derived in Section~\ref{sec:accretion}, we estimate that the mass accretion rate will reach $2\%$ Eddington at $t_{\rm obs} \approx 3-5$ years (or $t_{\rm obs}\approx 5-9$ years if instead the mass accretion rate more closely follows $\dot{M} \propto t^{-5/3}$) with an associated X-ray flux of $F_X \approx 1-5 \times 10^{-16} \ \rm erg \ cm^{-2} \ s^{-1}$. As this is beyond the detection limits of current X-ray facilities with reasonable exposure times, we conclude that continued X-ray monitoring of AT2022cmc is unlikely to detect a revival of the jetted emission, unless the transition to a ``low/hard'' state occurs at a larger fraction of the Eddington accretion rate. On the other hand, such a revival may be accompanied by a rapid rise in radio flux from the forward shock interaction of the jet with the ambient medium. Indeed, late-time radio rebrightening has been observed in a large fraction ($\sim 40\%$) of optically-selected TDEs \citep{Horesh2021,Cendes2023}, although the underlying mechanism governing the rebrightening, as well as whether the mechanism is the same for all TDEs, remains unclear. 

While jets have not been observed to turn back on for the existing sample of relativistic TDEs, late-time \textit{Chandra} observations of Sw J1644+57 have detected faint X-rays consistent with emission from the expanding forward shock \citep{Zauderer2013,Levan2016,Eftekhari2018}. X-ray observations of AT2022cmc at late times, coupled with an extrapolation of the radio SED into the X-ray regime, will probe such emission and enable constraints on the synchrotron cooling frequency. Finally, we note that the detection of quiescent X-ray emission from a pre-existing low-luminosity AGN is unlikely, given the distance to AT2022cmc and expected luminosities for low-luminosity AGN of $L_{X} \lesssim 10^{42} \rm erg \ s^{-1}$ \citep{Brusa2007}.


\section{Conclusions}

We have presented late-time X-ray observations of AT2022cmc, the fourth relativistic TDE discovered to date. Our observations reveal a drop in the X-ray emission at $t_{\rm rest} \approx 100$ days, which we attribute to the relativistic jet turning off, marking the third TDE for which such a transition is observed. Our main findings can be summarized as follows:

\begin{itemize}

\item We constrain the timescale of jet shut-off for AT2022cmc to $t_{\rm rest}\approx 98^{+38}_{-26}$ days. Compared to other jetted TDEs, the relativistic jet powering AT2022cmc shuts off a factor of two to four times earlier in the rest-frame, while the X-ray luminosities at shut-off (and hence Eddington luminosities) are comparable across all three events. The X-ray light curve prior to jet shut-off decays following $t^{-2.1}$, possibly indicative of a partial stellar disruption.

\item From the X-ray luminosity and timescale of jet shutoff, we parameterize the black hole mass in terms of the jet efficiency ($\eta_{\rm jet}$) and fraction of accreted matter ($f_{\rm in}$), and find that lower black hole masses of $\lesssim 10^5 \ \rm M_{\odot}$, where the disk and jet luminosities are comparable and the fraction of accreted mass is low, are favored. We find similar, albeit somewhat larger black hole masses of $\lesssim$ a few $\times 10^5 \ \rm M_{\odot}$ for Sw J1644+57 and Sw J2058+05, which we suggest may imply that jetted TDEs preferentially form in lower mass black hole systems than non-relativistic events (e.g., \citealt{Hammerstein2023a}) where the jet efficiency is low, and the corresponding disk efficiency is high \citep{Nicholl2022}.

\item We find that the overall accreted mass by $t_{\rm off} \approx 100$ days (rest-frame) is $\approx 0.1 \ M_{\odot}$, and the peak accretion rate is $\dot{M}_{p}\sim 4.6 \ \rm M_{\odot} \ yr^{-1}$. This is comparable to estimates of the total accreted mass for both Sw J1644+57 and Sw J2058+05. 

\item We predict that the jet associated with AT2022cmc will reach an accretion rate of 2\% Eddington and turn back on around $t_{\rm obs} \approx 3-9$ years depending on the mass fallback rate. However, the associated X-ray flux on this timescale of $F_X \approx 1-5 \times 10^{-16} \ \rm erg \ cm^{-2} \ s^{-1}$ precludes a detection with present-day X-ray facilities. 

\end{itemize}

While we attribute the sudden drop in X-ray emission for AT2022cmc to an accretion state transition, we note that several alternative models in which the jet becomes misaligned with our line of sight can reconcile the abrupt termination of X-rays in TDEs on timescales of $\sim$ years post-disruption. One such mechanism is the Bardeen-Petterson effect, in which the outer accretion disk is held tilted relative to the inner disk leading to a warped disk \citep{Bardeen1975}. Such a warp may induce a reorientation of the jet out of our line of sight \citep{Liska2021,Chatterjee2023}. Alternatively, the disk can be tilted due to stream self-intersections \citep{Curd2023}. As the self-intersection outflow weakens, the density contrast between the TDE stream and the disk increases resulting in a rapid realignment of the jet with the disk and an abrupt dropoff in the observed X-ray luminosity. 

As optical surveys contribute to an increasing rate of detections for jetted TDEs, X-ray monitoring on $\sim$ year timescales will enable constraints on the jet shut-off time for a larger sample of events and facilitate measurements of the SMBH mass function for relativistic TDEs. 

\appendix 
\label{sec:app}

\section{Constraining the Jet Shut-off Time in Sw J2058+05}

\begin{figure}
\center
\includegraphics[width=\columnwidth]{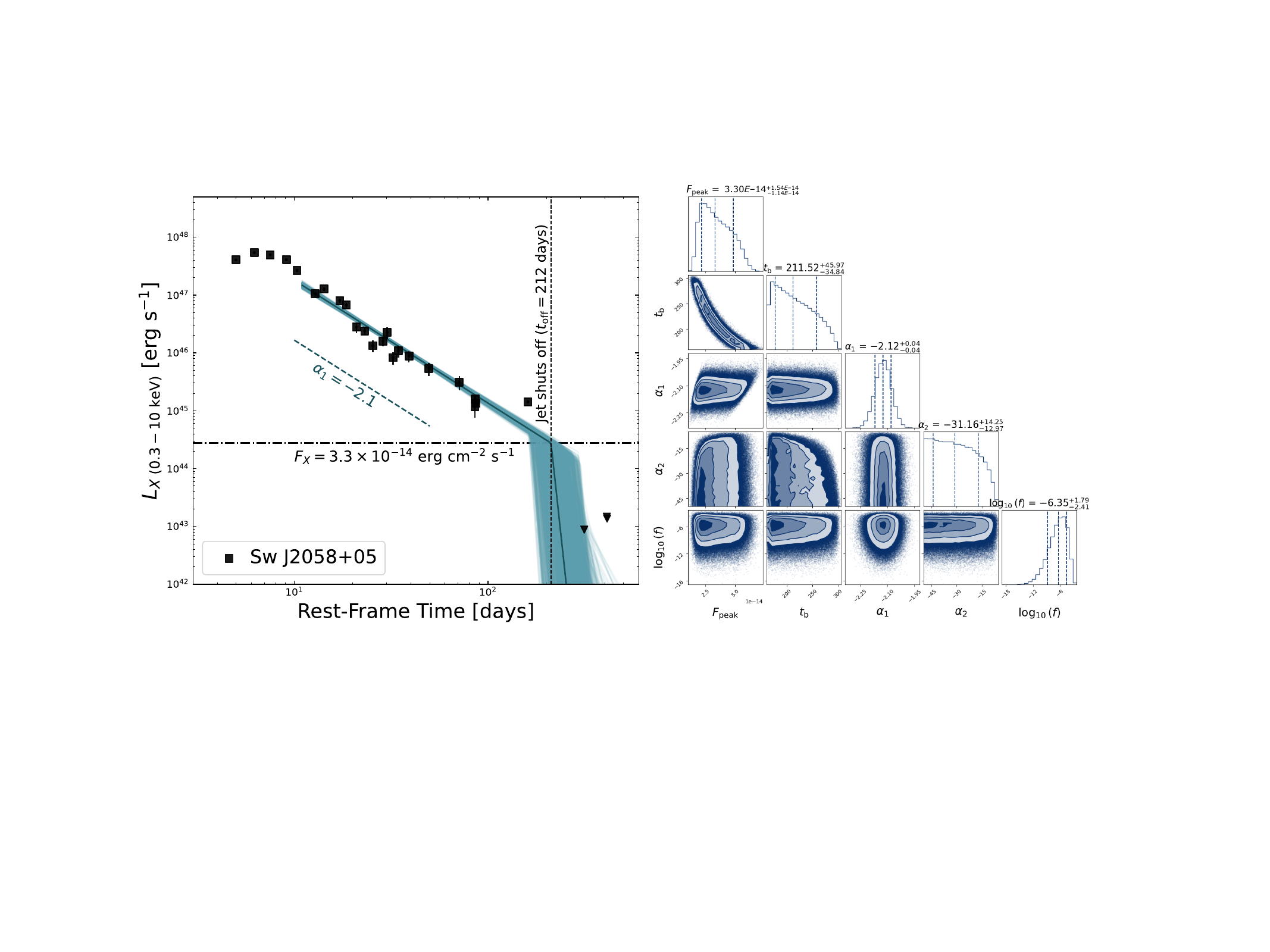}
\caption{\textbf{Left:} X-ray light curve of Sw J2058+05 (\citealt{Pasham2015}; black squares) with our best fit broken power-law model (dark blue curve)  which indicates a jet shut-off time of $t_{\rm rest}\approx 212^{+46}_{-35}$ days. The X-ray light curve prior to the jet shut-off decays following $t^{-2.1}$. \textbf{Right:} Results from our MCMC parameter estimation for a broken power law fit to the X-ray light curve of Sw J2058+05. Marginalized posterior distributions for each parameter are shown on the diagonal, where dashed lines indicate the median and $68\%$ confidence interval.}
\label{fig:swj2058}
\end{figure}

While the X-ray light curve of Sw J2058+05  (Figure~\ref{fig:swj2058}) exhibits a rapid decline following an earlier $\sim t^{-2}$ decay \citep{Pasham2015} attributable to the cessation of jet activity, as in the case of AT2022cmc, the precise timescale of jet-shutoff is not well-constrained given the sparse sampling of the X-ray light curve on this timescale. We therefore apply the above MCMC prescription to constrain the jet shut-off time for Sw J2058+05. We fit the light curve beginning at $t_{\rm rest} \approx 12$ days to account for the plateau in the X-ray light curve at $t\lesssim 12$ days. As in the case of AT2022cmc, we adopt log-uniform priors for $F_X$ and $t_{\rm off}$, and allow for $t_{\rm off}$ to span the full timescale of the X-ray light curve ($t\sim 0 - 410$ days). We use linearly uniform priors for the power law indices, with $-10 < \alpha_1 < 0$ and $-50 < \alpha_1 < -5$. We sample the posterior distributions using 1000 Markov chains and 30000 steps, and discard the first 3000 steps ($\approx 6 \times$ the autocorrelation length). We find an acceptance fraction of 0.36. From our MCMC analysis, we constrain the jet shut-off time in Sw J2058+05 to $t_{\rm rest} = 212^{+46}_{-35}$ days ($t_{\rm obs} \approx 460$ days).

\acknowledgments{The scientific results reported in this article are based on observations made by the \textit{Chandra X-ray Observatory} under program GO 24700310, PI: Eftekhari. T. E. is supported by NASA through the NASA Hubble Fellowship grant HST-HF2-51504.001-A awarded by the Space Telescope Science Institute, which is operated by the Association of Universities for Research in Astronomy, Inc., for NASA, under contract NAS5-26555. A. T. acknowledges support by NSF AST-2009884, AST-2107839, AST-1815304, AST-1911080, OAC-2031997, AST-2206471 and NASA 80NSSC22K0031 and 80NSSC21K1746 grants. K. D. A. acknowledges support provided by the NSF through award AST-2307668.}

\facilities{{\it Chandra X-ray Observatory}
}

\software{
\texttt{Astropy} \citep{astropy:2013, astropy:2018, astropy:2022},
\texttt{matplotlib} \citep{matplotlib},
\texttt{numpy} \citep{numpy}, 
\texttt{SAOImageDS9} \citep{DS9},
\texttt{scipy} \citep{scipy},
\texttt{CIAO} (v4.13) \citep{Fruscione2006},
}

\bibliographystyle{aasjournal}
\bibliography{references}

\begin{thebibliography}{}
\expandafter\ifx\csname natexlab\endcsname\relax\def\natexlab#1{#1}\fi
\providecommand{\url}[1]{\href{#1}{#1}}
\providecommand{\dodoi}[1]{doi:~\href{http://doi.org/#1}{\nolinkurl{#1}}}

\bibitem[{{Abramowicz} \& {Liu}(2012)}]{Abramowicz2012}
{Abramowicz}, M.~A., \& {Liu}, F.~K. 2012,
  \hypersetup{urlcolor=magenta}\href{https://dx.doi.org/10.1051/0004-6361/201220254}{\aap},
  \hypersetup{urlcolor=blue}\href{https://ui.adsabs.harvard.edu/abs/2012A&A...548A...3A}{548,
  A3}

\bibitem[{{Alexander} {et~al.}(2020){Alexander}, {van Velzen}, {Horesh}, \&
  {Zauderer}}]{Alexander2020}
{Alexander}, K.~D., {van Velzen}, S., {Horesh}, A., \& {Zauderer}, B.~A. 2020,
  \hypersetup{urlcolor=magenta}\href{https://dx.doi.org/10.1007/s11214-020-00702-w}{\ssr},
  \hypersetup{urlcolor=blue}\href{https://ui.adsabs.harvard.edu/abs/2020SSRv..216...81A}{216,
  81}

\bibitem[{{Andreoni} {et~al.}(2022{\natexlab{\hspace{0pt}a}}){Andreoni},
  {Coughlin}, {Perley}, {Yao}, {Lu}, {Cenko}, {Kumar}, {Anand}, {Ho},
  {Kasliwal}, {de Ugarte Postigo}, {Sagu{\'e}s-Carracedo}, {Schulze}, {Kann},
  {Kulkarni}, {Sollerman}, {Tanvir}, {Rest}, {Izzo}, {Somalwar}, {Kaplan},
  {Ahumada}, {Anupama}, {Auchettl}, {Barway}, {Bellm}, {Bhalerao}, {Bloom},
  {Bremer}, {Bulla}, {Burns}, {Campana}, {Chandra}, {Charalampopoulos},
  {Cooke}, {D'Elia}, {Das}, {Dobie}, {Ag{\"u}{\'\i} Fern{\'a}ndez}, {Freeburn},
  {Fremling}, {Gezari}, {Goode}, {Graham}, {Hammerstein}, {Karambelkar},
  {Kilpatrick}, {Kool}, {Krips}, {Laher}, {Leloudas}, {Levan}, {Lundquist},
  {Mahabal}, {Medford}, {Miller}, {M{\"o}ller}, {Mooley}, {Nayana}, {Nir},
  {Pang}, {Paraskeva}, {Perley}, {Petitpas}, {Pursiainen}, {Ravi},
  {Ridden-Harper}, {Riddle}, {Rigault}, {Rodriguez}, {Rusholme}, {Sharma},
  {Smith}, {Stein}, {Th{\"o}ne}, {Tohuvavohu}, {Valdes}, {van Roestel},
  {Vergani}, {Wang}, \& {Zhang}}]{Andreoni2022}
{Andreoni}, I., {Coughlin}, M.~W., {Perley}, D.~A., {et~al.}
  2022{\natexlab{\hspace{0pt}a}},
  \hypersetup{urlcolor=magenta}\href{https://dx.doi.org/10.1038/s41586-022-05465-8}{\nat},
  \hypersetup{urlcolor=blue}\href{https://ui.adsabs.harvard.edu/abs/2022Natur.612..430A}{612,
  430}

\bibitem[{{Andreoni} {et~al.}(2022{\natexlab{\hspace{0pt}b}}){Andreoni},
  {Coughlin}, {Ahumada}, {Kasliwal}, {Perley}, {Burns}, {Bulla}, {Cenko},
  {Anand}, \& {Kool}}]{Andreoni2022_gcn}
{Andreoni}, I., {Coughlin}, M., {Ahumada}, T., {et~al.}
  2022{\natexlab{\hspace{0pt}b}}, GRB Coordinates Network,
  \hypersetup{urlcolor=blue}\href{https://ui.adsabs.harvard.edu/abs/2022GCN.31590....1A}{31590,
  1}

\bibitem[{{Astropy Collaboration} {et~al.}(2013){Astropy Collaboration},
  {Robitaille}, {Tollerud}, {Greenfield}, {Droettboom}, {Bray}, {Aldcroft},
  {Davis}, {Ginsburg}, {Price-Whelan}, {Kerzendorf}, {Conley}, {Crighton},
  {Barbary}, {Muna}, {Ferguson}, {Grollier}, {Parikh}, {Nair}, {Unther},
  {Deil}, {Woillez}, {Conseil}, {Kramer}, {Turner}, {Singer}, {Fox}, {Weaver},
  {Zabalza}, {Edwards}, {Azalee Bostroem}, {Burke}, {Casey}, {Crawford},
  {Dencheva}, {Ely}, {Jenness}, {Labrie}, {Lim}, {Pierfederici}, {Pontzen},
  {Ptak}, {Refsdal}, {Servillat}, \& {Streicher}}]{astropy:2013}
{Astropy Collaboration}, {Robitaille}, T.~P., {Tollerud}, E.~J., {et~al.} 2013,
  \hypersetup{urlcolor=magenta}\href{https://dx.doi.org/10.1051/0004-6361/201322068}{\aap},
  \hypersetup{urlcolor=blue}\href{https://ui.adsabs.harvard.edu/abs/2013A&A...558A..33A}{558,
  A33}

\bibitem[{{Astropy Collaboration} {et~al.}(2018){Astropy Collaboration},
  {Price-Whelan}, {Sip{\H{o}}cz}, {G{\"u}nther}, {Lim}, {Crawford}, {Conseil},
  {Shupe}, {Craig}, {Dencheva}, {Ginsburg}, {VanderPlas}, {Bradley},
  {P{\'e}rez-Su{\'a}rez}, {de Val-Borro}, {Aldcroft}, {Cruz}, {Robitaille},
  {Tollerud}, {Ardelean}, {Babej}, {Bach}, {Bachetti}, {Bakanov}, {Bamford},
  {Barentsen}, {Barmby}, {Baumbach}, {Berry}, {Biscani}, {Boquien}, {Bostroem},
  {Bouma}, {Brammer}, {Bray}, {Breytenbach}, {Buddelmeijer}, {Burke},
  {Calderone}, {Cano Rodr{\'\i}guez}, {Cara}, {Cardoso}, {Cheedella}, {Copin},
  {Corrales}, {Crichton}, {D'Avella}, {Deil}, {Depagne}, {Dietrich}, {Donath},
  {Droettboom}, {Earl}, {Erben}, {Fabbro}, {Ferreira}, {Finethy}, {Fox},
  {Garrison}, {Gibbons}, {Goldstein}, {Gommers}, {Greco}, {Greenfield},
  {Groener}, {Grollier}, {Hagen}, {Hirst}, {Homeier}, {Horton}, {Hosseinzadeh},
  {Hu}, {Hunkeler}, {Ivezi{\'c}}, {Jain}, {Jenness}, {Kanarek}, {Kendrew},
  {Kern}, {Kerzendorf}, {Khvalko}, {King}, {Kirkby}, {Kulkarni}, {Kumar},
  {Lee}, {Lenz}, {Littlefair}, {Ma}, {Macleod}, {Mastropietro}, {McCully},
  {Montagnac}, {Morris}, {Mueller}, {Mumford}, {Muna}, {Murphy}, {Nelson},
  {Nguyen}, {Ninan}, {N{\"o}the}, {Ogaz}, {Oh}, {Parejko}, {Parley}, {Pascual},
  {Patil}, {Patil}, {Plunkett}, {Prochaska}, {Rastogi}, {Reddy Janga},
  {Sabater}, {Sakurikar}, {Seifert}, {Sherbert}, {Sherwood-Taylor}, {Shih},
  {Sick}, {Silbiger}, {Singanamalla}, {Singer}, {Sladen}, {Sooley},
  {Sornarajah}, {Streicher}, {Teuben}, {Thomas}, {Tremblay}, {Turner},
  {Terr{\'o}n}, {van Kerkwijk}, {de la Vega}, {Watkins}, {Weaver}, {Whitmore},
  {Woillez}, {Zabalza}, \& {Astropy Contributors}}]{astropy:2018}
{Astropy Collaboration}, {Price-Whelan}, A.~M., {Sip{\H{o}}cz}, B.~M., {et~al.}
  2018,
  \hypersetup{urlcolor=magenta}\href{https://dx.doi.org/10.3847/1538-3881/aabc4f}{\aj},
  \hypersetup{urlcolor=blue}\href{https://ui.adsabs.harvard.edu/abs/2018AJ....156..123A}{156,
  123}

\bibitem[{{Astropy Collaboration} {et~al.}(2022){Astropy Collaboration},
  {Price-Whelan}, {Lim}, {Earl}, {Starkman}, {Bradley}, {Shupe}, {Patil},
  {Corrales}, {Brasseur}, {N{\"o}the}, {Donath}, {Tollerud}, {Morris},
  {Ginsburg}, {Vaher}, {Weaver}, {Tocknell}, {Jamieson}, {van Kerkwijk},
  {Robitaille}, {Merry}, {Bachetti}, {G{\"u}nther}, {Aldcroft},
  {Alvarado-Montes}, {Archibald}, {B{\'o}di}, {Bapat}, {Barentsen},
  {Baz{\'a}n}, {Biswas}, {Boquien}, {Burke}, {Cara}, {Cara}, {Conroy},
  {Conseil}, {Craig}, {Cross}, {Cruz}, {D'Eugenio}, {Dencheva}, {Devillepoix},
  {Dietrich}, {Eigenbrot}, {Erben}, {Ferreira}, {Foreman-Mackey}, {Fox},
  {Freij}, {Garg}, {Geda}, {Glattly}, {Gondhalekar}, {Gordon}, {Grant},
  {Greenfield}, {Groener}, {Guest}, {Gurovich}, {Handberg}, {Hart},
  {Hatfield-Dodds}, {Homeier}, {Hosseinzadeh}, {Jenness}, {Jones}, {Joseph},
  {Kalmbach}, {Karamehmetoglu}, {Ka{\l}uszy{\'n}ski}, {Kelley}, {Kern},
  {Kerzendorf}, {Koch}, {Kulumani}, {Lee}, {Ly}, {Ma}, {MacBride}, {Maljaars},
  {Muna}, {Murphy}, {Norman}, {O'Steen}, {Oman}, {Pacifici}, {Pascual},
  {Pascual-Granado}, {Patil}, {Perren}, {Pickering}, {Rastogi}, {Roulston},
  {Ryan}, {Rykoff}, {Sabater}, {Sakurikar}, {Salgado}, {Sanghi}, {Saunders},
  {Savchenko}, {Schwardt}, {Seifert-Eckert}, {Shih}, {Jain}, {Shukla}, {Sick},
  {Simpson}, {Singanamalla}, {Singer}, {Singhal}, {Sinha}, {Sip{\H{o}}cz},
  {Spitler}, {Stansby}, {Streicher}, {{\v{S}}umak}, {Swinbank}, {Taranu},
  {Tewary}, {Tremblay}, {de Val-Borro}, {Van Kooten}, {Vasovi{\'c}}, {Verma},
  {de Miranda Cardoso}, {Williams}, {Wilson}, {Winkel}, {Wood-Vasey}, {Xue},
  {Yoachim}, {Zhang}, {Zonca}, \& {Astropy Project
  Contributors}}]{astropy:2022}
{Astropy Collaboration}, {Price-Whelan}, A.~M., {Lim}, P.~L., {et~al.} 2022,
  \hypersetup{urlcolor=magenta}\href{https://dx.doi.org/10.3847/1538-4357/ac7c74}{\apj},
  \hypersetup{urlcolor=blue}\href{https://ui.adsabs.harvard.edu/abs/2022ApJ...935..167A}{935,
  167}

\bibitem[{{Auchettl} {et~al.}(2018){Auchettl}, {Ramirez-Ruiz}, \&
  {Guillochon}}]{Auchettl2018}
{Auchettl}, K., {Ramirez-Ruiz}, E., \& {Guillochon}, J. 2018,
  \hypersetup{urlcolor=magenta}\href{https://dx.doi.org/10.3847/1538-4357/aa9b7c}{\apj},
  \hypersetup{urlcolor=blue}\href{https://ui.adsabs.harvard.edu/abs/2018ApJ...852...37A}{852,
  37}

\bibitem[{{Bardeen} \& {Petterson}(1975)}]{Bardeen1975}
{Bardeen}, J.~M., \& {Petterson}, J.~A. 1975,
  \hypersetup{urlcolor=magenta}\href{https://dx.doi.org/10.1086/181711}{\apjl},
  \hypersetup{urlcolor=blue}\href{https://ui.adsabs.harvard.edu/abs/1975ApJ...195L..65B}{195,
  L65}

\bibitem[{{Berger} {et~al.}(2012){Berger}, {Zauderer}, {Pooley}, {Soderberg},
  {Sari}, {Brunthaler}, \& {Bietenholz}}]{Berger2012}
{Berger}, E., {Zauderer}, A., {Pooley}, G.~G., {et~al.} 2012,
  \hypersetup{urlcolor=magenta}\href{https://dx.doi.org/10.1088/0004-637X/748/1/36}{\apj},
  \hypersetup{urlcolor=blue}\href{https://ui.adsabs.harvard.edu/abs/2012ApJ...748...36B}{748,
  36}

\bibitem[{{Blandford} \& {Znajek}(1977)}]{Blandford1977}
{Blandford}, R.~D., \& {Znajek}, R.~L. 1977,
  \hypersetup{urlcolor=magenta}\href{https://dx.doi.org/10.1093/mnras/179.3.433}{\mnras},
  \hypersetup{urlcolor=blue}\href{https://ui.adsabs.harvard.edu/abs/1977MNRAS.179..433B}{179,
  433}

\bibitem[{{Bloom} {et~al.}(2011){Bloom}, {Giannios}, {Metzger}, {Cenko},
  {Perley}, {Butler}, {Tanvir}, {Levan}, {O'Brien}, {Strubbe}, {De Colle},
  {Ramirez-Ruiz}, {Lee}, {Nayakshin}, {Quataert}, {King}, {Cucchiara},
  {Guillochon}, {Bower}, {Fruchter}, {Morgan}, \& {van der Horst}}]{Bloom2011}
{Bloom}, J.~S., {Giannios}, D., {Metzger}, B.~D., {et~al.} 2011,
  \hypersetup{urlcolor=magenta}\href{https://dx.doi.org/10.1126/science.1207150}{Science},
  \hypersetup{urlcolor=blue}\href{https://ui.adsabs.harvard.edu/abs/2011Sci...333..203B}{333,
  203}

\bibitem[{{Bonnerot} {et~al.}(2017){Bonnerot}, {Price}, {Lodato}, \&
  {Rossi}}]{Bonnerot2017}
{Bonnerot}, C., {Price}, D.~J., {Lodato}, G., \& {Rossi}, E.~M. 2017,
  \hypersetup{urlcolor=magenta}\href{https://dx.doi.org/10.1093/mnras/stx1210}{\mnras},
  \hypersetup{urlcolor=blue}\href{https://ui.adsabs.harvard.edu/abs/2017MNRAS.469.4879B}{469,
  4879}

\bibitem[{{Brown} {et~al.}(2015){Brown}, {Levan}, {Stanway}, {Tanvir}, {Cenko},
  {Berger}, {Chornock}, \& {Cucchiaria}}]{Brown2015}
{Brown}, G.~C., {Levan}, A.~J., {Stanway}, E.~R., {et~al.} 2015,
  \hypersetup{urlcolor=magenta}\href{https://dx.doi.org/10.1093/mnras/stv1520}{\mnras},
  \hypersetup{urlcolor=blue}\href{https://ui.adsabs.harvard.edu/abs/2015MNRAS.452.4297B}{452,
  4297}

\bibitem[{{Brusa} {et~al.}(2007){Brusa}, {Zamorani}, {Comastri}, {Hasinger},
  {Cappelluti}, {Civano}, {Finoguenov}, {Mainieri}, {Salvato}, {Vignali},
  {Elvis}, {Fiore}, {Gilli}, {Impey}, {Lilly}, {Mignoli}, {Silverman}, {Trump},
  {Urry}, {Bender}, {Capak}, {Huchra}, {Kneib}, {Koekemoer}, {Leauthaud},
  {Lehmann}, {Massey}, {Matute}, {McCarthy}, {McCracken}, {Rhodes}, {Scoville},
  {Taniguchi}, \& {Thompson}}]{Brusa2007}
{Brusa}, M., {Zamorani}, G., {Comastri}, A., {et~al.} 2007,
  \hypersetup{urlcolor=magenta}\href{https://dx.doi.org/10.1086/516575}{\apjs},
  \hypersetup{urlcolor=blue}\href{https://ui.adsabs.harvard.edu/abs/2007ApJS..172..353B}{172,
  353}

\bibitem[{{Burrows} {et~al.}(2011){Burrows}, {Kennea}, {Ghisellini}, {Mangano},
  {Zhang}, {Page}, {Eracleous}, {Romano}, {Sakamoto}, {Falcone}, {Osborne},
  {Campana}, {Beardmore}, {Breeveld}, {Chester}, {Corbet}, {Covino},
  {Cummings}, {D'Avanzo}, {D'Elia}, {Esposito}, {Evans}, {Fugazza}, {Gelbord},
  {Hiroi}, {Holland}, {Huang}, {Im}, {Israel}, {Jeon}, {Jeon}, {Jun}, {Kawai},
  {Kim}, {Krimm}, {Marshall}, {P. M{\'e}sz{\'a}ros}, {Negoro}, {Omodei},
  {Park}, {Perkins}, {Sugizaki}, {Sung}, {Tagliaferri}, {Troja}, {Ueda},
  {Urata}, {Usui}, {Antonelli}, {Barthelmy}, {Cusumano}, {Giommi}, {Melandri},
  {Perri}, {Racusin}, {Sbarufatti}, {Siegel}, \& {Gehrels}}]{Burrows2011}
{Burrows}, D.~N., {Kennea}, J.~A., {Ghisellini}, G., {et~al.} 2011,
  \hypersetup{urlcolor=magenta}\href{https://dx.doi.org/10.1038/nature10374}{\nat},
  \hypersetup{urlcolor=blue}\href{https://ui.adsabs.harvard.edu/abs/2011Natur.476..421B}{476,
  421}

\bibitem[{{Cendes} {et~al.}(2023){Cendes}, {Berger}, {Alexander}, {Chornock},
  {Margutti}, {Metzger}, {Wieringa}, {Bietenholz}, {Hajela}, {Laskar}, {Stroh},
  \& {Terreran}}]{Cendes2023}
{Cendes}, Y., {Berger}, E., {Alexander}, K.~D., {et~al.} 2023,
  \hypersetup{urlcolor=magenta}\href{https://dx.doi.org/10.48550/arXiv.2308.13595}{arXiv
  e-prints},
  \hypersetup{urlcolor=magenta}\href{https://arxiv.org/abs/2308.13595}{arXiv}{:}\hypersetup{urlcolor=blue}\href{https://ui.adsabs.harvard.edu/abs/2023arXiv230813595C}{2308.13595}

\bibitem[{{Cenko} {et~al.}(2012){Cenko}, {Krimm}, {Horesh}, {Rau}, {Frail},
  {Kennea}, {Levan}, {Holland}, {Butler}, {Quimby}, {Bloom}, {Filippenko},
  {Gal-Yam}, {Greiner}, {Kulkarni}, {Ofek}, {Olivares E.}, {Schady},
  {Silverman}, {Tanvir}, \& {Xu}}]{Cenko2012}
{Cenko}, S.~B., {Krimm}, H.~A., {Horesh}, A., {et~al.} 2012,
  \hypersetup{urlcolor=magenta}\href{https://dx.doi.org/10.1088/0004-637X/753/1/77}{\apj},
  \hypersetup{urlcolor=blue}\href{https://ui.adsabs.harvard.edu/abs/2012ApJ...753...77C}{753,
  77}

\bibitem[{{Chatterjee} {et~al.}(2023){Chatterjee}, {Liska}, {Tchekhovskoy}, \&
  {Markoff}}]{Chatterjee2023}
{Chatterjee}, K., {Liska}, M., {Tchekhovskoy}, A., \& {Markoff}, S. 2023,
  \hypersetup{urlcolor=magenta}\href{https://dx.doi.org/10.48550/arXiv.2311.00432}{arXiv
  e-prints},
  \hypersetup{urlcolor=magenta}\href{https://arxiv.org/abs/2311.00432}{arXiv}{:}\hypersetup{urlcolor=blue}\href{https://ui.adsabs.harvard.edu/abs/2023arXiv231100432C}{2311.00432}

\bibitem[{{Coughlin} \& {Nixon}(2019)}]{Coughlin2019}
{Coughlin}, E.~R., \& {Nixon}, C.~J. 2019,
  \hypersetup{urlcolor=magenta}\href{https://dx.doi.org/10.3847/2041-8213/ab412d}{\apjl},
  \hypersetup{urlcolor=blue}\href{https://ui.adsabs.harvard.edu/abs/2019ApJ...883L..17C}{883,
  L17}

\bibitem[{{Curd} {et~al.}(2023){Curd}, {Anantua}, {West}, \&
  {Duran}}]{Curd2023}
{Curd}, B., {Anantua}, R., {West}, H., \& {Duran}, J. 2023,
  \hypersetup{urlcolor=magenta}\href{https://dx.doi.org/10.48550/arXiv.2310.20592}{arXiv
  e-prints},
  \hypersetup{urlcolor=magenta}\href{https://arxiv.org/abs/2310.20592}{arXiv}{:}\hypersetup{urlcolor=blue}\href{https://ui.adsabs.harvard.edu/abs/2023arXiv231020592C}{2310.20592}

\bibitem[{{Curd} \& {Narayan}(2019)}]{Curd2019}
{Curd}, B., \& {Narayan}, R. 2019,
  \hypersetup{urlcolor=magenta}\href{https://dx.doi.org/10.1093/mnras/sty3134}{\mnras},
  \hypersetup{urlcolor=blue}\href{https://ui.adsabs.harvard.edu/abs/2019MNRAS.483..565C}{483,
  565}

\bibitem[{{Dai} {et~al.}(2021){Dai}, {Lodato}, \& {Cheng}}]{Dai2021}
{Dai}, J.~L., {Lodato}, G., \& {Cheng}, R. 2021,
  \hypersetup{urlcolor=magenta}\href{https://dx.doi.org/10.1007/s11214-020-00747-x}{\ssr},
  \hypersetup{urlcolor=blue}\href{https://ui.adsabs.harvard.edu/abs/2021SSRv..217...12D}{217,
  12}

\bibitem[{{Dai} {et~al.}(2018){Dai}, {McKinney}, {Roth}, {Ramirez-Ruiz}, \&
  {Miller}}]{Dai2018}
{Dai}, L., {McKinney}, J.~C., {Roth}, N., {Ramirez-Ruiz}, E., \& {Miller},
  M.~C. 2018,
  \hypersetup{urlcolor=magenta}\href{https://dx.doi.org/10.3847/2041-8213/aab429}{\apjl},
  \hypersetup{urlcolor=blue}\href{https://ui.adsabs.harvard.edu/abs/2018ApJ...859L..20D}{859,
  L20}

\bibitem[{{De Colle} {et~al.}(2012){De Colle}, {Guillochon}, {Naiman}, \&
  {Ramirez-Ruiz}}]{DeColle2012}
{De Colle}, F., {Guillochon}, J., {Naiman}, J., \& {Ramirez-Ruiz}, E. 2012,
  \hypersetup{urlcolor=magenta}\href{https://dx.doi.org/10.1088/0004-637X/760/2/103}{\apj},
  \hypersetup{urlcolor=blue}\href{https://ui.adsabs.harvard.edu/abs/2012ApJ...760..103D}{760,
  103}

\bibitem[{{De Colle} \& {Lu}(2020)}]{DeColle2020}
{De Colle}, F., \& {Lu}, W. 2020,
  \hypersetup{urlcolor=magenta}\href{https://dx.doi.org/10.1016/j.newar.2020.101538}{\nar},
  \hypersetup{urlcolor=blue}\href{https://ui.adsabs.harvard.edu/abs/2020NewAR..8901538D}{89,
  101538}

\bibitem[{{Eftekhari} {et~al.}(2018){Eftekhari}, {Berger}, {Zauderer},
  {Margutti}, \& {Alexander}}]{Eftekhari2018}
{Eftekhari}, T., {Berger}, E., {Zauderer}, B.~A., {Margutti}, R., \&
  {Alexander}, K.~D. 2018,
  \hypersetup{urlcolor=magenta}\href{https://dx.doi.org/10.3847/1538-4357/aaa8e0}{\apj},
  \hypersetup{urlcolor=blue}\href{https://ui.adsabs.harvard.edu/abs/2018ApJ...854...86E}{854,
  86}

\bibitem[{{Fender} {et~al.}(2004){Fender}, {Belloni}, \& {Gallo}}]{Fender2004}
{Fender}, R.~P., {Belloni}, T.~M., \& {Gallo}, E. 2004,
  \hypersetup{urlcolor=magenta}\href{https://dx.doi.org/10.1111/j.1365-2966.2004.08384.x}{\mnras},
  \hypersetup{urlcolor=blue}\href{https://ui.adsabs.harvard.edu/abs/2004MNRAS.355.1105F}{355,
  1105}

\bibitem[{{Foreman-Mackey} {et~al.}(2013){Foreman-Mackey}, {Hogg}, {Lang}, \&
  {Goodman}}]{Foreman-Mackey2013}
{Foreman-Mackey}, D., {Hogg}, D.~W., {Lang}, D., \& {Goodman}, J. 2013,
  \hypersetup{urlcolor=magenta}\href{https://dx.doi.org/10.1086/670067}{\pasp},
  \hypersetup{urlcolor=blue}\href{https://ui.adsabs.harvard.edu/abs/2013PASP..125..306F}{125,
  306}

\bibitem[{{Franchini} {et~al.}(2016){Franchini}, {Lodato}, \&
  {Facchini}}]{Franchini2016}
{Franchini}, A., {Lodato}, G., \& {Facchini}, S. 2016,
  \hypersetup{urlcolor=magenta}\href{https://dx.doi.org/10.1093/mnras/stv2417}{\mnras},
  \hypersetup{urlcolor=blue}\href{https://ui.adsabs.harvard.edu/abs/2016MNRAS.455.1946F}{455,
  1946}

\bibitem[{{Fruscione} {et~al.}(2006){Fruscione}, {McDowell}, {Allen},
  {Brickhouse}, {Burke}, {Davis}, {Durham}, {Elvis}, {Galle}, {Harris},
  {Huenemoerder}, {Houck}, {Ishibashi}, {Karovska}, {Nicastro}, {Noble},
  {Nowak}, {Primini}, {Siemiginowska}, {Smith}, \& {Wise}}]{Fruscione2006}
{Fruscione}, A., {McDowell}, J.~C., {Allen}, G.~E., {et~al.} 2006, in Society
  of Photo-Optical Instrumentation Engineers (SPIE) Conference Series, Vol.
  6270, Society of Photo-Optical Instrumentation Engineers (SPIE) Conference
  Series, ed. D.~R. {Silva} \& R.~E. {Doxsey}, 62701V

\bibitem[{{Gehrels}(1986)}]{Gehrels1986}
{Gehrels}, N. 1986,
  \hypersetup{urlcolor=magenta}\href{https://dx.doi.org/10.1086/164079}{\apj},
  \hypersetup{urlcolor=blue}\href{https://ui.adsabs.harvard.edu/abs/1986ApJ...303..336G}{303,
  336}

\bibitem[{Gelman {et~al.}(1996)Gelman, Roberts, \& Gilks}]{Gelman1996}
Gelman, A., Roberts, G.~O., \& Gilks, W.~R. 1996, in Bayesian Statistics, ed.
  J.~M. Bernardo, J.~O. Berger, A.~P. Dawid, \& A.~F.~M. Smith (Oxford
  University Press, Oxford)

\bibitem[{{Ghisellini} {et~al.}(2010){Ghisellini}, {Tavecchio}, {Foschini},
  {Ghirlanda}, {Maraschi}, \& {Celotti}}]{Ghisellini2010}
{Ghisellini}, G., {Tavecchio}, F., {Foschini}, L., {et~al.} 2010,
  \hypersetup{urlcolor=magenta}\href{https://dx.doi.org/10.1111/j.1365-2966.2009.15898.x}{\mnras},
  \hypersetup{urlcolor=blue}\href{https://ui.adsabs.harvard.edu/abs/2010MNRAS.402..497G}{402,
  497}

\bibitem[{{Giannios} \& {Metzger}(2011)}]{Giannios2011}
{Giannios}, D., \& {Metzger}, B.~D. 2011,
  \hypersetup{urlcolor=magenta}\href{https://dx.doi.org/10.1111/j.1365-2966.2011.19188.x}{\mnras},
  \hypersetup{urlcolor=blue}\href{https://ui.adsabs.harvard.edu/abs/2011MNRAS.416.2102G}{416,
  2102}

\bibitem[{{Guillochon} \& {McCourt}(2017)}]{Guillochon2017}
{Guillochon}, J., \& {McCourt}, M. 2017,
  \hypersetup{urlcolor=magenta}\href{https://dx.doi.org/10.3847/2041-8213/834/2/L19}{\apjl},
  \hypersetup{urlcolor=blue}\href{https://ui.adsabs.harvard.edu/abs/2017ApJ...834L..19G}{834,
  L19}

\bibitem[{{Guillochon} \& {Ramirez-Ruiz}(2013)}]{Guillochon2013}
{Guillochon}, J., \& {Ramirez-Ruiz}, E. 2013,
  \hypersetup{urlcolor=magenta}\href{https://dx.doi.org/10.1088/0004-637X/767/1/25}{\apj},
  \hypersetup{urlcolor=blue}\href{https://ui.adsabs.harvard.edu/abs/2013ApJ...767...25G}{767,
  25}

\bibitem[{{Hammerstein} {et~al.}(2023){Hammerstein}, {van Velzen}, {Gezari},
  {Cenko}, {Yao}, {Ward}, {Frederick}, {Villanueva}, {Somalwar}, {Graham},
  {Kulkarni}, {Stern}, {Andreoni}, {Bellm}, {Dekany}, {Dhawan}, {Drake},
  {Fremling}, {Gatkine}, {Groom}, {Ho}, {Kasliwal}, {Karambelkar}, {Kool},
  {Masci}, {Medford}, {Perley}, {Purdum}, {van Roestel}, {Sharma}, {Sollerman},
  {Taggart}, \& {Yan}}]{Hammerstein2023a}
{Hammerstein}, E., {van Velzen}, S., {Gezari}, S., {et~al.} 2023,
  \hypersetup{urlcolor=magenta}\href{https://dx.doi.org/10.3847/1538-4357/aca283}{\apj},
  \hypersetup{urlcolor=blue}\href{https://ui.adsabs.harvard.edu/abs/2023ApJ...942....9H}{942,
  9}

\bibitem[{{H{\"a}ring} \& {Rix}(2004)}]{Haring2004}
{H{\"a}ring}, N., \& {Rix}, H.-W. 2004,
  \hypersetup{urlcolor=magenta}\href{https://dx.doi.org/10.1086/383567}{\apjl},
  \hypersetup{urlcolor=blue}\href{https://ui.adsabs.harvard.edu/abs/2004ApJ...604L..89H}{604,
  L89}

\bibitem[{{Harris} {et~al.}(2020){Harris}, {Millman}, {van der Walt},
  {Gommers}, {Virtanen}, {Cournapeau}, {Wieser}, {Taylor}, {Berg}, {Smith},
  {Kern}, {Picus}, {Hoyer}, {van Kerkwijk}, {Brett}, {Haldane}, {del R{\'\i}o},
  {Wiebe}, {Peterson}, {G{\'e}rard-Marchant}, {Sheppard}, {Reddy}, {Weckesser},
  {Abbasi}, {Gohlke}, \& {Oliphant}}]{numpy}
{Harris}, C.~R., {Millman}, K.~J., {van der Walt}, S.~J., {et~al.} 2020,
  \hypersetup{urlcolor=magenta}\href{https://dx.doi.org/10.1038/s41586-020-2649-2}{\nat},
  \hypersetup{urlcolor=blue}\href{https://ui.adsabs.harvard.edu/abs/2020Natur.585..357H}{585,
  357}

\bibitem[{{Helsel}(2005)}]{Helsel2005}
{Helsel}, D.~R. 2005,
  \hypersetup{urlcolor=magenta}\href{https://dx.doi.org/10.1021/es053368a}{Environmental
  Science and Technology},
  \hypersetup{urlcolor=blue}\href{https://ui.adsabs.harvard.edu/abs/2005EnST...39..419H}{39,
  419A}

\bibitem[{{Horesh} {et~al.}(2021{\natexlab{\hspace{0pt}a}}){Horesh}, {Cenko},
  \& {Arcavi}}]{Horesh2021b}
{Horesh}, A., {Cenko}, S.~B., \& {Arcavi}, I. 2021{\natexlab{\hspace{0pt}a}},
  \hypersetup{urlcolor=magenta}\href{https://dx.doi.org/10.1038/s41550-021-01300-8}{Nature
  Astronomy},
  \hypersetup{urlcolor=blue}\href{https://ui.adsabs.harvard.edu/abs/2021NatAs...5..491H}{5,
  491}

\bibitem[{{Horesh} {et~al.}(2021{\natexlab{\hspace{0pt}b}}){Horesh}, {Sfaradi},
  {Fender}, {Green}, {Williams}, \& {Bright}}]{Horesh2021}
{Horesh}, A., {Sfaradi}, I., {Fender}, R., {et~al.}
  2021{\natexlab{\hspace{0pt}b}},
  \hypersetup{urlcolor=magenta}\href{https://dx.doi.org/10.3847/2041-8213/ac25fe}{\apjl},
  \hypersetup{urlcolor=blue}\href{https://ui.adsabs.harvard.edu/abs/2021ApJ...920L...5H}{920,
  L5}

\bibitem[{{Hunter}(2007)}]{matplotlib}
{Hunter}, J.~D. 2007,
  \hypersetup{urlcolor=magenta}\href{https://dx.doi.org/10.1109/MCSE.2007.55}{Computing
  in Science and Engineering},
  \hypersetup{urlcolor=blue}\href{https://ui.adsabs.harvard.edu/abs/2007CSE.....9...90H}{9,
  90}

\bibitem[{{Inoue} {et~al.}(2017){Inoue}, {Doi}, {Tanaka}, {Sikora}, \&
  {Madejski}}]{Inoue2017}
{Inoue}, Y., {Doi}, A., {Tanaka}, Y.~T., {Sikora}, M., \& {Madejski}, G.~M.
  2017,
  \hypersetup{urlcolor=magenta}\href{https://dx.doi.org/10.3847/1538-4357/aa6b57}{\apj},
  \hypersetup{urlcolor=blue}\href{https://ui.adsabs.harvard.edu/abs/2017ApJ...840...46I}{840,
  46}

\bibitem[{{Jiang} {et~al.}(2019){Jiang}, {Stone}, \& {Davis}}]{Jiang2019}
{Jiang}, Y.-F., {Stone}, J.~M., \& {Davis}, S.~W. 2019,
  \hypersetup{urlcolor=magenta}\href{https://dx.doi.org/10.3847/1538-4357/ab29ff}{\apj},
  \hypersetup{urlcolor=blue}\href{https://ui.adsabs.harvard.edu/abs/2019ApJ...880...67J}{880,
  67}

\bibitem[{{Joye} \& {Mandel}(2003)}]{DS9}
{Joye}, W.~A., \& {Mandel}, E. 2003, in Astronomical Society of the Pacific
  Conference Series, Vol. 295, Astronomical Data Analysis Software and Systems
  XII, ed. H.~E. {Payne}, R.~I. {Jedrzejewski}, \& R.~N. {Hook}, 489

\bibitem[{{Kalberla} {et~al.}(2005){Kalberla}, {Burton}, {Hartmann}, {Arnal},
  {Bajaja}, {Morras}, \& {P{\"o}ppel}}]{Kalberla2005}
{Kalberla}, P.~M.~W., {Burton}, W.~B., {Hartmann}, D., {et~al.} 2005,
  \hypersetup{urlcolor=magenta}\href{https://dx.doi.org/10.1051/0004-6361:20041864}{\aap},
  \hypersetup{urlcolor=blue}\href{https://ui.adsabs.harvard.edu/abs/2005A&A...440..775K}{440,
  775}

\bibitem[{{Kelley} {et~al.}(2014){Kelley}, {Tchekhovskoy}, \&
  {Narayan}}]{Kelley2014}
{Kelley}, L.~Z., {Tchekhovskoy}, A., \& {Narayan}, R. 2014,
  \hypersetup{urlcolor=magenta}\href{https://dx.doi.org/10.1093/mnras/stu2041}{\mnras},
  \hypersetup{urlcolor=blue}\href{https://ui.adsabs.harvard.edu/abs/2014MNRAS.445.3919K}{445,
  3919}

\bibitem[{{Krolik} \& {Piran}(2011)}]{Krolik2011}
{Krolik}, J.~H., \& {Piran}, T. 2011,
  \hypersetup{urlcolor=magenta}\href{https://dx.doi.org/10.1088/0004-637X/743/2/134}{\apj},
  \hypersetup{urlcolor=blue}\href{https://ui.adsabs.harvard.edu/abs/2011ApJ...743..134K}{743,
  134}

\bibitem[{{Krolik} \& {Piran}(2012)}]{Krolik2012}
{Krolik}, J.~H., \& {Piran}, T. 2012,
  \hypersetup{urlcolor=magenta}\href{https://dx.doi.org/10.1088/0004-637X/749/1/92}{\apj},
  \hypersetup{urlcolor=blue}\href{https://ui.adsabs.harvard.edu/abs/2012ApJ...749...92K}{749,
  92}

\bibitem[{{Laskar} {et~al.}(2014){Laskar}, {Berger}, {Tanvir}, {Zauderer},
  {Margutti}, {Levan}, {Perley}, {Fong}, {Wiersema}, {Menten}, \&
  {Hrudkova}}]{Laskar2014}
{Laskar}, T., {Berger}, E., {Tanvir}, N., {et~al.} 2014,
  \hypersetup{urlcolor=magenta}\href{https://dx.doi.org/10.1088/0004-637X/781/1/1}{\apj},
  \hypersetup{urlcolor=blue}\href{https://ui.adsabs.harvard.edu/abs/2014ApJ...781....1L}{781,
  1}

\bibitem[{Lawless(2002)}]{Lawless2002}
Lawless, J.~F. 2002, Statistical Models and Methods for Lifetime Data
  (\hypersetup{urlcolor=magenta}\href{https://doi.org/10.1002/9781118033005}{Wiley})

\bibitem[{{Levan} {et~al.}(2011){Levan}, {Tanvir}, {Cenko}, {Perley},
  {Wiersema}, {Bloom}, {Fruchter}, {de Ugarte Postigo}, {O'Brien}, {Butler},
  {van der Horst}, {Leloudas}, {Morgan}, {Misra}, {Bower}, {Farihi},
  {Tunnicliffe}, {Modjaz}, {Silverman}, {Hjorth}, {Th{\"o}ne}, {Cucchiara},
  {Cer{\'o}n}, {Castro-Tirado}, {Arnold}, {Bremer}, {Brodie}, {Carroll},
  {Cooper}, {Curran}, {Cutri}, {Ehle}, {Forbes}, {Fynbo}, {Gorosabel},
  {Graham}, {Hoffman}, {Guziy}, {Jakobsson}, {Kamble}, {Kerr}, {Kasliwal},
  {Kouveliotou}, {Kocevski}, {Law}, {Nugent}, {Ofek}, {Poznanski}, {Quimby},
  {Rol}, {Romanowsky}, {S{\'a}nchez-Ram{\'\i}rez}, {Schulze}, {Singh}, {van
  Spaandonk}, {Starling}, {Strom}, {Tello}, {Vaduvescu}, {Wheatley}, {Wijers},
  {Winters}, \& {Xu}}]{Levan2011}
{Levan}, A.~J., {Tanvir}, N.~R., {Cenko}, S.~B., {et~al.} 2011,
  \hypersetup{urlcolor=magenta}\href{https://dx.doi.org/10.1126/science.1207143}{Science},
  \hypersetup{urlcolor=blue}\href{https://ui.adsabs.harvard.edu/abs/2011Sci...333..199L}{333,
  199}

\bibitem[{{Levan} {et~al.}(2016){Levan}, {Tanvir}, {Brown}, {Metzger}, {Page},
  {Cenko}, {O'Brien}, {Lyman}, {Wiersema}, {Stanway}, {Fruchter}, {Perley}, \&
  {Bloom}}]{Levan2016}
{Levan}, A.~J., {Tanvir}, N.~R., {Brown}, G.~C., {et~al.} 2016,
  \hypersetup{urlcolor=magenta}\href{https://dx.doi.org/10.3847/0004-637X/819/1/51}{\apj},
  \hypersetup{urlcolor=blue}\href{https://ui.adsabs.harvard.edu/abs/2016ApJ...819...51L}{819,
  51}

\bibitem[{{Liska} {et~al.}(2021){Liska}, {Hesp}, {Tchekhovskoy}, {Ingram}, {van
  der Klis}, {Markoff}, \& {Van Moer}}]{Liska2021}
{Liska}, M., {Hesp}, C., {Tchekhovskoy}, A., {et~al.} 2021,
  \hypersetup{urlcolor=magenta}\href{https://dx.doi.org/10.1093/mnras/staa099}{\mnras},
  \hypersetup{urlcolor=blue}\href{https://ui.adsabs.harvard.edu/abs/2021MNRAS.507..983L}{507,
  983}

\bibitem[{{Lu} \& {Bonnerot}(2020)}]{Lu2020}
{Lu}, W., \& {Bonnerot}, C. 2020,
  \hypersetup{urlcolor=magenta}\href{https://dx.doi.org/10.1093/mnras/stz3405}{\mnras},
  \hypersetup{urlcolor=blue}\href{https://ui.adsabs.harvard.edu/abs/2020MNRAS.492..686L}{492,
  686}

\bibitem[{{Lu} {et~al.}(2023){Lu}, {Matsumoto}, \& {Matzner}}]{Lu2023}
{Lu}, W., {Matsumoto}, T., \& {Matzner}, C.~D. 2023,
  \hypersetup{urlcolor=magenta}\href{https://dx.doi.org/10.48550/arXiv.2310.15336}{arXiv
  e-prints},
  \hypersetup{urlcolor=magenta}\href{https://arxiv.org/abs/2310.15336}{arXiv}{:}\hypersetup{urlcolor=blue}\href{https://ui.adsabs.harvard.edu/abs/2023arXiv231015336L}{2310.15336}

\bibitem[{{Maccarone}(2003)}]{Maccarone2003}
{Maccarone}, T.~J. 2003,
  \hypersetup{urlcolor=magenta}\href{https://dx.doi.org/10.1051/0004-6361:20031146}{\aap},
  \hypersetup{urlcolor=blue}\href{https://ui.adsabs.harvard.edu/abs/2003A&A...409..697M}{409,
  697}

\bibitem[{{Mangano} {et~al.}(2016){Mangano}, {Burrows}, {Sbarufatti}, \&
  {Cannizzo}}]{Mangano2016}
{Mangano}, V., {Burrows}, D.~N., {Sbarufatti}, B., \& {Cannizzo}, J.~K. 2016,
  \hypersetup{urlcolor=magenta}\href{https://dx.doi.org/10.3847/0004-637X/817/2/103}{\apj},
  \hypersetup{urlcolor=blue}\href{https://ui.adsabs.harvard.edu/abs/2016ApJ...817..103M}{817,
  103}

\bibitem[{{McConnell} \& {Ma}(2013)}]{McConnell2013}
{McConnell}, N.~J., \& {Ma}, C.-P. 2013,
  \hypersetup{urlcolor=magenta}\href{https://dx.doi.org/10.1088/0004-637X/764/2/184}{\apj},
  \hypersetup{urlcolor=blue}\href{https://ui.adsabs.harvard.edu/abs/2013ApJ...764..184M}{764,
  184}

\bibitem[{{Metzger} \& {Stone}(2016)}]{Metzger2016}
{Metzger}, B.~D., \& {Stone}, N.~C. 2016,
  \hypersetup{urlcolor=magenta}\href{https://dx.doi.org/10.1093/mnras/stw1394}{\mnras},
  \hypersetup{urlcolor=blue}\href{https://ui.adsabs.harvard.edu/abs/2016MNRAS.461..948M}{461,
  948}

\bibitem[{{Mockler} {et~al.}(2019){Mockler}, {Guillochon}, \&
  {Ramirez-Ruiz}}]{Mockler2019}
{Mockler}, B., {Guillochon}, J., \& {Ramirez-Ruiz}, E. 2019,
  \hypersetup{urlcolor=magenta}\href{https://dx.doi.org/10.3847/1538-4357/ab010f}{\apj},
  \hypersetup{urlcolor=blue}\href{https://ui.adsabs.harvard.edu/abs/2019ApJ...872..151M}{872,
  151}

\bibitem[{{Nagar} {et~al.}(2005){Nagar}, {Falcke}, \& {Wilson}}]{Nagar2005}
{Nagar}, N.~M., {Falcke}, H., \& {Wilson}, A.~S. 2005,
  \hypersetup{urlcolor=magenta}\href{https://dx.doi.org/10.1051/0004-6361:20042277}{\aap},
  \hypersetup{urlcolor=blue}\href{https://ui.adsabs.harvard.edu/abs/2005A&A...435..521N}{435,
  521}

\bibitem[{{Nicholl} {et~al.}(2022){Nicholl}, {Lanning}, {Ramsden}, {Mockler},
  {Lawrence}, {Short}, \& {Ridley}}]{Nicholl2022}
{Nicholl}, M., {Lanning}, D., {Ramsden}, P., {et~al.} 2022,
  \hypersetup{urlcolor=magenta}\href{https://dx.doi.org/10.1093/mnras/stac2206}{\mnras},
  \hypersetup{urlcolor=blue}\href{https://ui.adsabs.harvard.edu/abs/2022MNRAS.515.5604N}{515,
  5604}

\bibitem[{{Novikov} \& {Thorne}(1973)}]{Novikov1973}
{Novikov}, I.~D., \& {Thorne}, K.~S. 1973, in Black Holes (Les Astres Occlus),
  343--450

\bibitem[{{Pasham} {et~al.}(2015){Pasham}, {Cenko}, {Levan}, {Bower}, {Horesh},
  {Brown}, {Dolan}, {Wiersema}, {Filippenko}, {Fruchter}, {Greiner}, {O'Brien},
  {Page}, {Rau}, \& {Tanvir}}]{Pasham2015}
{Pasham}, D.~R., {Cenko}, S.~B., {Levan}, A.~J., {et~al.} 2015,
  \hypersetup{urlcolor=magenta}\href{https://dx.doi.org/10.1088/0004-637X/805/1/68}{\apj},
  \hypersetup{urlcolor=blue}\href{https://ui.adsabs.harvard.edu/abs/2015ApJ...805...68P}{805,
  68}

\bibitem[{{Pasham} {et~al.}(2023){Pasham}, {Lucchini}, {Laskar}, {Gompertz},
  {Srivastav}, {Nicholl}, {Smartt}, {Miller-Jones}, {Alexander}, {Fender},
  {Smith}, {Fulton}, {Dewangan}, {Gendreau}, {Coughlin}, {Rhodes}, {Horesh},
  {van Velzen}, {Sfaradi}, {Guolo}, {Castro Segura}, {Aamer}, {Anderson},
  {Arcavi}, {Brennan}, {Chambers}, {Charalampopoulos}, {Chen}, {Clocchiatti},
  {de Boer}, {Dennefeld}, {Ferrara}, {Galbany}, {Gao}, {Gillanders}, {Goodwin},
  {Gromadzki}, {Huber}, {Jonker}, {Joshi}, {Kara}, {Killestein}, {Kosec},
  {Kocevski}, {Leloudas}, {Lin}, {Margutti}, {Mattila}, {Moore},
  {M{\"u}ller-Bravo}, {Ngeow}, {Oates}, {Onori}, {Pan}, {Perez-Torres}, {Rani},
  {Remillard}, {Ridley}, {Schulze}, {Sheng}, {Shingles}, {Smith}, {Steiner},
  {Wainscoat}, {Wevers}, \& {Yang}}]{Pasham2023}
{Pasham}, D.~R., {Lucchini}, M., {Laskar}, T., {et~al.} 2023,
  \hypersetup{urlcolor=magenta}\href{https://dx.doi.org/10.1038/s41550-022-01820-x}{Nature
  Astronomy},
  \hypersetup{urlcolor=blue}\href{https://ui.adsabs.harvard.edu/abs/2023NatAs...7...88P}{7,
  88}

\bibitem[{{Piran} {et~al.}(2015){Piran}, {S{\k{a}}dowski}, \&
  {Tchekhovskoy}}]{Piran2015}
{Piran}, T., {S{\k{a}}dowski}, A., \& {Tchekhovskoy}, A. 2015,
  \hypersetup{urlcolor=magenta}\href{https://dx.doi.org/10.1093/mnras/stv1547}{\mnras},
  \hypersetup{urlcolor=blue}\href{https://ui.adsabs.harvard.edu/abs/2015MNRAS.453..157P}{453,
  157}

\bibitem[{{Planck Collaboration} {et~al.}(2020){Planck Collaboration},
  {Aghanim}, {Akrami}, {Ashdown}, {Aumont}, {Baccigalupi}, {Ballardini},
  {Banday}, {Barreiro}, {Bartolo}, {Basak}, {Battye}, {Benabed}, {Bernard},
  {Bersanelli}, {Bielewicz}, {Bock}, {Bond}, {Borrill}, {Bouchet}, {Boulanger},
  {Bucher}, {Burigana}, {Butler}, {Calabrese}, {Cardoso}, {Carron},
  {Challinor}, {Chiang}, {Chluba}, {Colombo}, {Combet}, {Contreras}, {Crill},
  {Cuttaia}, {de Bernardis}, {de Zotti}, {Delabrouille}, {Delouis}, {Di
  Valentino}, {Diego}, {Dor{\'e}}, {Douspis}, {Ducout}, {Dupac}, {Dusini},
  {Efstathiou}, {Elsner}, {En{\ss}lin}, {Eriksen}, {Fantaye}, {Farhang},
  {Fergusson}, {Fernandez-Cobos}, {Finelli}, {Forastieri}, {Frailis},
  {Fraisse}, {Franceschi}, {Frolov}, {Galeotta}, {Galli}, {Ganga},
  {G{\'e}nova-Santos}, {Gerbino}, {Ghosh}, {Gonz{\'a}lez-Nuevo}, {G{\'o}rski},
  {Gratton}, {Gruppuso}, {Gudmundsson}, {Hamann}, {Handley}, {Hansen},
  {Herranz}, {Hildebrandt}, {Hivon}, {Huang}, {Jaffe}, {Jones}, {Karakci},
  {Keih{\"a}nen}, {Keskitalo}, {Kiiveri}, {Kim}, {Kisner}, {Knox},
  {Krachmalnicoff}, {Kunz}, {Kurki-Suonio}, {Lagache}, {Lamarre}, {Lasenby},
  {Lattanzi}, {Lawrence}, {Le Jeune}, {Lemos}, {Lesgourgues}, {Levrier},
  {Lewis}, {Liguori}, {Lilje}, {Lilley}, {Lindholm}, {L{\'o}pez-Caniego},
  {Lubin}, {Ma}, {Mac{\'\i}as-P{\'e}rez}, {Maggio}, {Maino}, {Mandolesi},
  {Mangilli}, {Marcos-Caballero}, {Maris}, {Martin}, {Martinelli},
  {Mart{\'\i}nez-Gonz{\'a}lez}, {Matarrese}, {Mauri}, {McEwen}, {Meinhold},
  {Melchiorri}, {Mennella}, {Migliaccio}, {Millea}, {Mitra},
  {Miville-Desch{\^e}nes}, {Molinari}, {Montier}, {Morgante}, {Moss}, {Natoli},
  {N{\o}rgaard-Nielsen}, {Pagano}, {Paoletti}, {Partridge}, {Patanchon},
  {Peiris}, {Perrotta}, {Pettorino}, {Piacentini}, {Polastri}, {Polenta},
  {Puget}, {Rachen}, {Reinecke}, {Remazeilles}, {Renzi}, {Rocha}, {Rosset},
  {Roudier}, {Rubi{\~n}o-Mart{\'\i}n}, {Ruiz-Granados}, {Salvati}, {Sandri},
  {Savelainen}, {Scott}, {Shellard}, {Sirignano}, {Sirri}, {Spencer},
  {Sunyaev}, {Suur-Uski}, {Tauber}, {Tavagnacco}, {Tenti}, {Toffolatti},
  {Tomasi}, {Trombetti}, {Valenziano}, {Valiviita}, {Van Tent}, {Vibert},
  {Vielva}, {Villa}, {Vittorio}, {Wandelt}, {Wehus}, {White}, {White},
  {Zacchei}, \& {Zonca}}]{Planck2020}
{Planck Collaboration}, {Aghanim}, N., {Akrami}, Y., {et~al.} 2020,
  \hypersetup{urlcolor=magenta}\href{https://dx.doi.org/10.1051/0004-6361/201833910}{\aap},
  \hypersetup{urlcolor=blue}\href{https://ui.adsabs.harvard.edu/abs/2020A&A...641A...6P}{641,
  A6}

\bibitem[{{Rees}(1988)}]{Rees1988}
{Rees}, M.~J. 1988,
  \hypersetup{urlcolor=magenta}\href{https://dx.doi.org/10.1038/333523a0}{\nat},
  \hypersetup{urlcolor=blue}\href{https://ui.adsabs.harvard.edu/abs/1988Natur.333..523R}{333,
  523}

\bibitem[{{Reis} {et~al.}(2012){Reis}, {Miller}, {Reynolds}, {G{\"u}ltekin},
  {Maitra}, {King}, \& {Strohmayer}}]{Reis2012}
{Reis}, R.~C., {Miller}, J.~M., {Reynolds}, M.~T., {et~al.} 2012,
  \hypersetup{urlcolor=magenta}\href{https://dx.doi.org/10.1126/science.1223940}{Science},
  \hypersetup{urlcolor=blue}\href{https://ui.adsabs.harvard.edu/abs/2012Sci...337..949R}{337,
  949}

\bibitem[{{Rhodes} {et~al.}(2023){Rhodes}, {Bright}, {Fender}, {Sfaradi},
  {Green}, {Horesh}, {Mooley}, {Pasham}, {Smartt}, {Titterington}, {van der
  Horst}, \& {Williams}}]{Rhodes2023}
{Rhodes}, L., {Bright}, J.~S., {Fender}, R., {et~al.} 2023,
  \hypersetup{urlcolor=magenta}\href{https://dx.doi.org/10.1093/mnras/stad344}{\mnras},
  \hypersetup{urlcolor=blue}\href{https://ui.adsabs.harvard.edu/abs/2023MNRAS.521..389R}{521,
  389}

\bibitem[{{Russell} {et~al.}(2011){Russell}, {Miller-Jones}, {Maccarone},
  {Yang}, {Fender}, \& {Lewis}}]{Russell2011}
{Russell}, D.~M., {Miller-Jones}, J.~C.~A., {Maccarone}, T.~J., {et~al.} 2011,
  \hypersetup{urlcolor=magenta}\href{https://dx.doi.org/10.1088/2041-8205/739/1/L19}{\apjl},
  \hypersetup{urlcolor=blue}\href{https://ui.adsabs.harvard.edu/abs/2011ApJ...739L..19R}{739,
  L19}

\bibitem[{{Ryu} {et~al.}(2020){Ryu}, {Krolik}, \& {Piran}}]{Ryu2020}
{Ryu}, T., {Krolik}, J., \& {Piran}, T. 2020,
  \hypersetup{urlcolor=magenta}\href{https://dx.doi.org/10.3847/1538-4357/abbf4d}{\apj},
  \hypersetup{urlcolor=blue}\href{https://ui.adsabs.harvard.edu/abs/2020ApJ...904...73R}{904,
  73}

\bibitem[{{Sari} {et~al.}(1999){Sari}, {Piran}, \& {Halpern}}]{Sari1999}
{Sari}, R., {Piran}, T., \& {Halpern}, J.~P. 1999,
  \hypersetup{urlcolor=magenta}\href{https://dx.doi.org/10.1086/312109}{\apjl},
  \hypersetup{urlcolor=blue}\href{https://ui.adsabs.harvard.edu/abs/1999ApJ...519L..17S}{519,
  L17}

\bibitem[{{Sato} {et~al.}(2024){Sato}, {Murase}, {Bhattacharya}, {Carpio},
  {Mukhopadhyay}, \& {Zhang}}]{Sato2024}
{Sato}, Y., {Murase}, K., {Bhattacharya}, M., {et~al.} 2024,
  \hypersetup{urlcolor=magenta}\href{https://dx.doi.org/10.48550/arXiv.2404.13326}{arXiv
  e-prints},
  \hypersetup{urlcolor=magenta}\href{https://arxiv.org/abs/2404.13326}{arXiv}{:}\hypersetup{urlcolor=blue}\href{https://ui.adsabs.harvard.edu/abs/2024arXiv240413326S}{2404.13326}

\bibitem[{{Saxton} {et~al.}(2012){Saxton}, {Soria}, {Wu}, \&
  {Kuin}}]{Saxton2012}
{Saxton}, C.~J., {Soria}, R., {Wu}, K., \& {Kuin}, N. P.~M. 2012,
  \hypersetup{urlcolor=magenta}\href{https://dx.doi.org/10.1111/j.1365-2966.2012.20739.x}{\mnras},
  \hypersetup{urlcolor=blue}\href{https://ui.adsabs.harvard.edu/abs/2012MNRAS.422.1625S}{422,
  1625}

\bibitem[{{Sfaradi} {et~al.}(2024){Sfaradi}, {Beniamini}, {Horesh}, {Piran},
  {Bright}, {Rhodes}, {Williams}, {Fender}, {Leung}, {Murphy}, \&
  {Green}}]{Sfaradi2024}
{Sfaradi}, I., {Beniamini}, P., {Horesh}, A., {et~al.} 2024,
  \hypersetup{urlcolor=magenta}\href{https://dx.doi.org/10.1093/mnras/stad3717}{\mnras},
  \hypersetup{urlcolor=blue}\href{https://ui.adsabs.harvard.edu/abs/2024MNRAS.527.7672S}{527,
  7672}

\bibitem[{{Shakura} \& {Sunyaev}(1973)}]{Shakura1973}
{Shakura}, N.~I., \& {Sunyaev}, R.~A. 1973, \aap,
  \hypersetup{urlcolor=blue}\href{https://ui.adsabs.harvard.edu/abs/1973A&A....24..337S}{24,
  337}

\bibitem[{{Shen} \& {Matzner}(2014)}]{Shen2014}
{Shen}, R.-F., \& {Matzner}, C.~D. 2014,
  \hypersetup{urlcolor=magenta}\href{https://dx.doi.org/10.1088/0004-637X/784/2/87}{\apj},
  \hypersetup{urlcolor=blue}\href{https://ui.adsabs.harvard.edu/abs/2014ApJ...784...87S}{784,
  87}

\bibitem[{{S{\k{a}}dowski} \& {Narayan}(2016)}]{Sadowski2016}
{S{\k{a}}dowski}, A., \& {Narayan}, R. 2016,
  \hypersetup{urlcolor=magenta}\href{https://dx.doi.org/10.1093/mnras/stv2941}{\mnras},
  \hypersetup{urlcolor=blue}\href{https://ui.adsabs.harvard.edu/abs/2016MNRAS.456.3929S}{456,
  3929}

\bibitem[{{Somalwar} {et~al.}(2023){Somalwar}, {Ravi}, {Dong}, {Hammerstein},
  {Hallinan}, {Law}, {Miller}, {Myers}, {Yao}, {Dekany}, {Graham}, {Groom},
  {Purdum}, \& {Wold}}]{Somolwar2023}
{Somalwar}, J.~J., {Ravi}, V., {Dong}, D.~Z., {et~al.} 2023,
  \hypersetup{urlcolor=magenta}\href{https://dx.doi.org/10.48550/arXiv.2310.03791}{arXiv
  e-prints},
  \hypersetup{urlcolor=magenta}\href{https://arxiv.org/abs/2310.03791}{arXiv}{:}\hypersetup{urlcolor=blue}\href{https://ui.adsabs.harvard.edu/abs/2023arXiv231003791S}{2310.03791}

\bibitem[{Stone \& Loeb(2012)}]{Stone2012}
Stone, N., \& Loeb, A. 2012,
  \hypersetup{urlcolor=magenta}\href{https://dx.doi.org/10.1103/physrevlett.108.061302}{Physical
  Review Letters},
  \hypersetup{urlcolor=blue}\href{https://ui.adsabs.harvard.edu/#abs/2012S}{108}

\bibitem[{{Stone} {et~al.}(2013){Stone}, {Sari}, \& {Loeb}}]{Stone2013}
{Stone}, N., {Sari}, R., \& {Loeb}, A. 2013,
  \hypersetup{urlcolor=magenta}\href{https://dx.doi.org/10.1093/mnras/stt1270}{\mnras},
  \hypersetup{urlcolor=blue}\href{https://ui.adsabs.harvard.edu/abs/2013MNRAS.435.1809S}{435,
  1809}

\bibitem[{{Stone} {et~al.}(2020){Stone}, {Vasiliev}, {Kesden}, {Rossi},
  {Perets}, \& {Amaro-Seoane}}]{Stone2020}
{Stone}, N.~C., {Vasiliev}, E., {Kesden}, M., {et~al.} 2020,
  \hypersetup{urlcolor=magenta}\href{https://dx.doi.org/10.1007/s11214-020-00651-4}{\ssr},
  \hypersetup{urlcolor=blue}\href{https://ui.adsabs.harvard.edu/abs/2020SSRv..216...35S}{216,
  35}

\bibitem[{{Strubbe} \& {Quataert}(2009)}]{Strubbe2009}
{Strubbe}, L.~E., \& {Quataert}, E. 2009,
  \hypersetup{urlcolor=magenta}\href{https://dx.doi.org/10.1111/j.1365-2966.2009.15599.x}{\mnras},
  \hypersetup{urlcolor=blue}\href{https://ui.adsabs.harvard.edu/abs/2009MNRAS.400.2070S}{400,
  2070}

\bibitem[{Su {et~al.}(2016)Su, Liu, \& Zhang}]{Su2016}
Su, R., Liu, X., \& Zhang, Z. 2016,
  \hypersetup{urlcolor=magenta}\href{https://dx.doi.org/10.1007/s10509-016-2990-y}{Astrophysics
  and Space Science},
  \hypersetup{urlcolor=blue}\href{https://ui.adsabs.harvard.edu/#abs/2016S}{362}

\bibitem[{{Sun} {et~al.}(2015){Sun}, {Zhang}, \& {Li}}]{Sun2015}
{Sun}, H., {Zhang}, B., \& {Li}, Z. 2015,
  \hypersetup{urlcolor=magenta}\href{https://dx.doi.org/10.1088/0004-637X/812/1/33}{\apj},
  \hypersetup{urlcolor=blue}\href{https://ui.adsabs.harvard.edu/abs/2015ApJ...812...33S}{812,
  33}

\bibitem[{{Tanvir} {et~al.}(2022){Tanvir}, {de Ugarte Postigo}, {Izzo},
  {Vergani}, {D'Elia}, {Campana}, {Perley}, {Wiersema}, {Levan}, {Kann},
  {Rossi}, {Della Valle}, \& {Stargate Consortium}}]{Tanvir2022}
{Tanvir}, N.~R., {de Ugarte Postigo}, A., {Izzo}, L., {et~al.} 2022, GRB
  Coordinates Network,
  \hypersetup{urlcolor=blue}\href{https://ui.adsabs.harvard.edu/abs/2022GCN.31602....1T}{31602,
  1}

\bibitem[{{Tchekhovskoy} {et~al.}(2014){Tchekhovskoy}, {Metzger}, {Giannios},
  \& {Kelley}}]{Tchekhovskoy2014}
{Tchekhovskoy}, A., {Metzger}, B.~D., {Giannios}, D., \& {Kelley}, L.~Z. 2014,
  \hypersetup{urlcolor=magenta}\href{https://dx.doi.org/10.1093/mnras/stt2085}{\mnras},
  \hypersetup{urlcolor=blue}\href{https://ui.adsabs.harvard.edu/abs/2014MNRAS.437.2744T}{437,
  2744}

\bibitem[{{Teboul} \& {Metzger}(2023)}]{Teboul2023}
{Teboul}, O., \& {Metzger}, B.~D. 2023,
  \hypersetup{urlcolor=magenta}\href{https://dx.doi.org/10.3847/2041-8213/ad0037}{\apjl},
  \hypersetup{urlcolor=blue}\href{https://ui.adsabs.harvard.edu/abs/2023ApJ...957L...9T}{957,
  L9}

\bibitem[{{Ulmer}(1999)}]{Ulmer1999}
{Ulmer}, A. 1999,
  \hypersetup{urlcolor=magenta}\href{https://dx.doi.org/10.1086/306909}{\apj},
  \hypersetup{urlcolor=blue}\href{https://ui.adsabs.harvard.edu/abs/1999ApJ...514..180U}{514,
  180}

\bibitem[{Virtanen {et~al.}(2020)Virtanen, Gommers, Oliphant, Haberland, Reddy,
  Cournapeau, Burovski, Peterson, Weckesser, Bright, {van der Walt}, Brett,
  Wilson, Millman, Mayorov, Nelson, Jones, Kern, Larson, Carey, Polat, Feng,
  Moore, {VanderPlas}, Laxalde, Perktold, Cimrman, Henriksen, Quintero, Harris,
  Archibald, Ribeiro, Pedregosa, {van Mulbregt}, \& {SciPy 1.0
  Contributors}}]{scipy}
Virtanen, P., Gommers, R., Oliphant, T.~E., {et~al.} 2020,
  \hypersetup{urlcolor=magenta}\href{https://dx.doi.org/10.1038/s41592-019-0686-2}{Nature
  Methods}, \hypersetup{urlcolor=blue}\href{https://rdcu.be/b08Wh}{17, 261}

\bibitem[{{Wang} {et~al.}(2018){Wang}, {Zhang}, {Liang}, {Lu}, {Lin}, {Li}, \&
  {Li}}]{Wang2018}
{Wang}, X.-G., {Zhang}, B., {Liang}, E.-W., {et~al.} 2018,
  \hypersetup{urlcolor=magenta}\href{https://dx.doi.org/10.3847/1538-4357/aabc13}{\apj},
  \hypersetup{urlcolor=blue}\href{https://ui.adsabs.harvard.edu/abs/2018ApJ...859..160W}{859,
  160}

\bibitem[{{Yao} {et~al.}(2024){Yao}, {Lu}, {Harrison}, {Kulkarni}, {Gezari},
  {Guolo}, {Cenko}, \& {Ho}}]{Yao2024}
{Yao}, Y., {Lu}, W., {Harrison}, F., {et~al.} 2024,
  \hypersetup{urlcolor=magenta}\href{https://dx.doi.org/10.3847/1538-4357/ad2b6b}{\apj},
  \hypersetup{urlcolor=blue}\href{https://ui.adsabs.harvard.edu/abs/2024ApJ...965...39Y}{965,
  39}

\bibitem[{{Zanazzi} \& {Lai}(2019)}]{Zanazzi2019}
{Zanazzi}, J.~J., \& {Lai}, D. 2019,
  \hypersetup{urlcolor=magenta}\href{https://dx.doi.org/10.1093/mnras/stz1610}{\mnras},
  \hypersetup{urlcolor=blue}\href{https://ui.adsabs.harvard.edu/abs/2019MNRAS.487.4965Z}{487,
  4965}

\bibitem[{{Zauderer} {et~al.}(2013){Zauderer}, {Berger}, {Margutti}, {Pooley},
  {Sari}, {Soderberg}, {Brunthaler}, \& {Bietenholz}}]{Zauderer2013}
{Zauderer}, B.~A., {Berger}, E., {Margutti}, R., {et~al.} 2013,
  \hypersetup{urlcolor=magenta}\href{https://dx.doi.org/10.1088/0004-637X/767/2/152}{\apj},
  \hypersetup{urlcolor=blue}\href{https://ui.adsabs.harvard.edu/abs/2013ApJ...767..152Z}{767,
  152}

\bibitem[{{Zauderer} {et~al.}(2011){Zauderer}, {Berger}, {Soderberg}, {Loeb},
  {Narayan}, {Frail}, {Petitpas}, {Brunthaler}, {Chornock}, {Carpenter},
  {Pooley}, {Mooley}, {Kulkarni}, {Margutti}, {Fox}, {Nakar}, {Patel},
  {Volgenau}, {Culverhouse}, {Bietenholz}, {Rupen}, {Max-Moerbeck}, {Readhead},
  {Richards}, {Shepherd}, {Storm}, \& {Hull}}]{Zauderer2011}
{Zauderer}, B.~A., {Berger}, E., {Soderberg}, A.~M., {et~al.} 2011,
  \hypersetup{urlcolor=magenta}\href{https://dx.doi.org/10.1038/nature10366}{\nat},
  \hypersetup{urlcolor=blue}\href{https://ui.adsabs.harvard.edu/abs/2011Natur.476..425Z}{476,
  425}

\end{thebibliography}

\end{document}